\documentclass[12pt,twoside,english]{iopart}
\usepackage[T1]{fontenc}
\usepackage[latin9]{inputenc}
\usepackage[letterpaper]{geometry}
\usepackage{float}
\usepackage{units}
\usepackage{textcomp}
\usepackage{amsthm}
\usepackage{amstext}
\usepackage{graphicx}
\usepackage{amssymb}

\makeatletter
\usepackage{iopams}
\usepackage{setstack}
\theoremstyle{plain}
\newtheorem{thm}{Theorem}
\theoremstyle{definition}
\newtheorem{defn}[thm]{Definition}



\usepackage{babel}

\makeatother

\usepackage{babel}

\usepackage[all]{xy}

\makeatother

\usepackage{babel}

\begin{document}
\global\long\global\long\def\rr{\mathbb{R}}
 \global\long\global\long\def\cc{\mathbb{C}}
\global\long\global\long\def\tg{\tilde{\Gamma}}
 \global\long\def\todo#1{{\bf TODO: #1\/}}
\global\long\global\long\def\vin{V_{\Gamma}^{in}\left(k\right)}
\global\long\global\long\def\vout{V_{\Gamma}^{out}\left(k\right)}
\global\long\global\long\def\vtin{V_{\tg}^{in}}
\global\long\global\long\def\vtout{V_{\tg}^{out}}
\global\long\global\long\def\vv{\mathcal{V}}
 \global\long\def\ee{\mathcal{E}}
 \global\long\def\leads{\mathcal{L}}
\global\long\global\long\def\lgr{L_{\nicefrac{\tg}{R}}}
 \global\long\global\long\def\lgi{L_{\tg}^{(i)}}
\global\long\global\long\def\lgri{L_{\nicefrac{\tg}{R}}^{(i)}}

\title{Scattering from isospectral quantum graphs}

\author{R.~Band$^{1}$}

\address{$^{\text{1}}$Department of Physics of Complex Systems, The Weizmann
Institute of Science, Rehovot 76100, Israel}

\author{A.~Sawicki$^{1,\,2}$}

\address{$^{\text{2}}$Center for Theoretical Physics, Polish Academy of Sciences
Al. Lotnik\'ow 32/46, 02-668 Warszawa, Poland}

\author{U.~Smilansky$^{1,\,3}$}

\address{$^{\text{3}}$Cardiff School of Mathematics and WIMCS, Cardiff University,
Senghennydd Road, Cardiff CF24 4AG, UK}
\begin{abstract}
Quantum graphs can be extended to scattering systems when they are
connected by leads to infinity. It is shown that for certain extensions,
the scattering matrices of isospectral graphs are conjugate to each
other and their poles distributions are therefore identical. The scattering
matrices are studied using a recently developed isospectral theory
\cite{BPB09,PB09}. At the same time, the scattering approach offers
a new insight on the mentioned isospectral construction.
\end{abstract}
\maketitle

\section{Introduction}

The investigation of spectral inverse problems goes back to the famous
question of Marc Kac, 'Can one hear the shape of a drum?' \cite{Kac66}.
This question can be rephrased as 'does the Laplacian on every planar
domain with Dirichlet boundary conditions have a unique spectrum?'.
Ever since the time when Kac posed this fascinating question, physicists
and mathematicians alike have attacked the problem from various angles.
Attempts were made both to reconstruct the shape of an object from
its spectrum and to find different objects that are isospectral, i.e.,
have the same spectrum. In 1985, Sunada presented a method for constructing
isospectral Riemannian manifolds \cite{Sunada85}. Over the years,
several pairs of isospectral objects were found, but these were not
planar domains, and therefore did not serve as an exact answer to
Kac's question. In 1992, by applying an extension of Sunada's theorem,
Gordon, Webb and Wolpert were able to finally answer Kac's question
as it related to drums, presenting the first pair of isospectral two-dimensional
planar domains \cite{GWW92a, GWW92b}. At the same time, the research of inverse
spectral problems extended to include the examination of scattering
data. Examples of objects that share the same scattering information
were found both for finite area \cite{Berard92,Zelditch92} and infinite
 area \cite{GuiZwo97,Brooks02isoscatteringon} Riemann surfaces. The
search for isospectral and isoscattering examples went further than
expected by the original question of Kac - it now includes various
approaches and a wealth of objects to consider, ranging from Riemannian
manifolds to discrete graphs. The interested reader can find more
about the field in the reviews \cite{MR2179793,Gordon09,MR1705572}
and the references within.

The main result in the field of isospectrality and isoscattering of
quantum graphs is that of Gutkin and Smilansky \cite{GS01}, where
they use the trace formula to solve the inverse spectral problem of
a general quantum graph. They show that under certain conditions a
quantum graph can be recovered either from the spectrum of its Laplacian
or from the overall phase of its scattering matrix. The necessary
conditions include the graph being simple and its edges having rationally
independent lengths. When these conditions are not satisfied, isospectral
quantum graphs indeed arise. Examples of isospectral and isoscattering
quantum graphs appear in \cite{Roth83,vonBelow00,Oren08,GS01,SS05,BSS06,MR959124,Kostrykin:403589,KosSch06,Boman200558,0305-4470-35-1-309}.
A recent work \cite{BPB09,PB09} presents a new method for constructing
isospectral objects. This method is a generalization of Sunada's theorem
of isospectrality \cite{Sunada85}, as it relaxes its hypothesis.

The present paper makes use of the above mentioned isospectral theory
in order to investigate isoscattering problems and relate them to
the isospectral research. We were motivated by \cite{OSTH05}, in
which scattering from the exterior of isospectral domains in $\mathbb{R}^{2}$
is discussed. Okada et al. conjecture in this paper that one may distinguish
isospectral drums by measuring sound scattered by the drums. In other
words, in spite of the fact that the two domains are isospectral,
the authors suggest that when looked from the exterior, the corresponding
scattering matrices have different poles distributions, i.e., the
domains are not isoscattering. The results that we obtain in this
manuscript for quantum graphs shed a new light on the above conjecture.

The paper is arranged as follows. In the present section we shortly
review quantum graphs in the context of isospectrality and scattering
matrices. We then bring a basic example which we use throughout  the
paper to demonstrate the obtained results. Section \ref{sec:isospectrality_and_scattering}
develops a theory which connects scattering matrices to isospectrality
and the next section applies it to a few examples. We end by indicating
the link to \cite{OSTH05}, summarizing and suggesting future research
directions.

\subsection{Quantum graphs}

\label{sec:quantum_graphs}

Let $\Gamma=\left(\vv,\,\ee\right)$ be a finite graph which consists
of $\left|\vv\right|$ vertices connected by
$\left|\mathcal{E}\right|$ edges. Each edge $e\in\ee$ is a one
dimensional segment of finite length $L_{e}$ with a coordinate
$x_{e}\in\left[0,L_{e}\right]$ and this makes $\Gamma$ a metric
graph. The metric graph becomes quantum, when we supply it with a
differential operator. In the present paper we choose our operator
to be the free Schr\"{o}dinger operator and denote it by $\Delta$.
This is merely the one-dimensional Laplacian which equals
$-\frac{{\rm d}^{2}}{{\rm d}x_{e}^{2}}$ on each of the edges
$e\in\ee$, and its domain is
$\bigoplus_{e\in\ee}H^{2}\left(e\right)$. The coupling between the
edges is introduced by supplementing vertex conditions at the
vertices. There are many choices of vertex conditions which render
the resulting operator self-adjoint, and the most common ones are
the Neumann vertex conditions, described below.

Let $v\in\vv$, and $\ee_{v}$ the set of edges incident to $v$.
A function $f$ on $\Gamma$ obeys Neumann vertex conditions at $v$
if and only if
\begin{enumerate}
\item $f$ is continuous at $v$, i.e., \[
\forall e_{1},e_{2}\in\ee_{v}\,;\,\, f_{e_{1}}\left(v\right)=f_{e_{2}}\left(v\right).\]

\item The sum of outgoing derivatives of $f$ at the vertex $v$ equals
zero, i.e.,\[
\sum_{e\in\ee_{v}}\frac{{\rm \textrm{d}f}}{\textrm{d}x_{e}}\left(v\right)=0.\]

\end{enumerate}
The spectrum of a graph all of whose vertices obey Neumann conditions
is discrete, non-negative and unbounded.

Other typical conditions, which are called Dirichlet, set to zero
the values of the function at the vertex,\[
\forall e\in\ee_{v}\,;\,\, f_{e}\left(v\right)=0,\]
and do not put any requirement on the derivatives.

In general, the vertex conditions at a certain vertex $v\in\vv$ can
be described in one of two ways:
\begin{enumerate}
\item Stating linear equations for the values and the derivatives of the
function at the vertex. Specifically, we use $\left|\ee_{v}\right|$
equations for the $2\left|\ee_{v}\right|$ variables: $\left\{ f_{e}\left(v\right)\right\} _{e\in\ee_{v}}$
and $\left\{ \frac{{\rm \textrm{d}f}}{\textrm{d}x_{e}}\left(v\right)\right\} _{e\in\ee_{v}}$.
\item Representing the function on each edge $e\in\ee_{v}$ as a linear
combination of two exponents, \[
f_{e}\left(x_{e}\right)=a_{e}^{in}\exp\left(-ikx_{e}\right)+a_{e}^{out}\exp\left(ikx_{e}\right),\]
and dictating linear relations between the vectors of amplitudes $\vec{a}^{in}\in\cc^{\left|\ee_{v}\right|}$
and $\vec{a}^{out}\in\cc^{\left|\ee_{v}\right|}$. The vertex conditions
are then expressed using an a priori given unitary matrix $\sigma_{v}$,
as \[
\vec{a}^{out}=\sigma_{v}\vec{a}^{in}.\]

\end{enumerate}
A more detailed description on these characterizations of vertex conditions
and the relations between them appears in \cite{GS06,Kuchment04}.

\subsection{Isospectral graphs and their transplantation\label{sec:isospectrality_and_transplantation}}

The recent papers \cite{BPB09,PB09} present a new isospectral construction
method which is based on basic elements of representation theory and
can be applied for any geometric object. We bring here the relevant
aspects of the theory as it applies to quantum graphs \footnote{The interested reader is referred to appendix A in \cite{BPB09} for a short review of the Algebra used in this section.}.

Let $\Gamma$ be a graph which obeys a certain symmetry group $G$.
This means that each element of $G$ is a graph automorphism which
preserves both the lengths of the edges and the vertex conditions.
Denote by $\Phi_{\Gamma}\left(k\right)$ the eigenspace of the Laplacian
on $\Gamma$ with eigenvalue $k^{2}$. The action of $G$ on $\Gamma$
induces an action of $G$ on $\Phi_{\Gamma}\left(k\right)$, by\[
\forall x\in\Gamma,\,\forall g\in G,\,\forall f\in\Phi_{\Gamma}\left(k\right);\,\,\left(g\cdot f\right)\left(x\right)=f\left(g^{-1}x\right).\]
Since $\Phi_{\Gamma}\left(k\right)$ is closed under the action of
$G$, we have that $\Phi_{\Gamma}\left(k\right)$ is a carrier space
of a certain representation of $G$. Let $R$ be another representation
of $G$ with some abstract carrier space $V^{R}$. We consider all
linear transformations $\varphi:V^{R}\rightarrow\Phi_{\Gamma}\left(k\right)$
that respect the action of the group $G$, i.e., \[
\forall g\in G,\,\forall v\in V^{R}\,;\,\varphi\left(g\cdot v\right)=g\cdot\varphi\left(v\right).\]
These linear transformations are called intertwiners and they form
a vector space which is denoted by $\mathrm{Hom}_{\cc G}\left(V^{R},\Phi_{\Gamma}\left(k\right)\right)$.
When $R$ is an irreducible representation, each such an intertwiner
$\varphi$ is an embedding of $V^{R}$ in the eigenspace $\Phi_{\Gamma}\left(k\right)$.
The dimension of $\mathrm{Hom}_{\cc G}\left(V^{R},\Phi_{\Gamma}\left(k\right)\right)$
in this case equals the number of copies of $V^{R}$ that are contained
within $\Phi_{\Gamma}\left(k\right)$.

It was shown in \cite{BPB09,PB09} that there exists a graph, denoted
by $\nicefrac{\Gamma}{R}$, which obeys \begin{equation}
\forall k\quad\Phi\left(k\right)\cong\mathrm{Hom}_{\mathbb{C}G}\left(V^{R},\Phi_{\Gamma}\left(k\right)\right).\label{eq:quotietgreigen}\end{equation}
These papers supply a theorem which gives a sufficient condition for
two quotient graphs $\nicefrac{\Gamma}{R_{1}}$ and $\nicefrac{\Gamma}{R_{2}}$
to be isospectral, where $R_{1},\, R_{2}$ are representations of
two symmetry groups $H_{1},\, H_{2}\leq G$ of $\Gamma$. The condition
is $\textrm{Ind}_{H_{1}}^{G}R_{1}\cong\textrm{Ind}_{H_{2}}^{G}R_{2}$.

An important element of the isospectral research is the concept of
\emph{transplantation}. This is a transformation between the eigenspace
of one object to the eigenspace of its isospectral partner, with the
same eigenvalue. Specifically, if $\Gamma_{1},\,\Gamma_{2}$ are isospectral
graphs, then a transplantation is an isomorphism $T:\Phi_{\Gamma_{1}}\left(k\right)\overset{\cong}{\longrightarrow}\Phi_{\Gamma_{2}}\left(k\right)$.
A transplantation is therefore quite a useful tool to prove the isospectrality
of objects. The isospectral construction method described in \cite{BPB09,PB09}
guarantees a corresponding transplantation. In general, isospectral
objects consist of some elementary \textquotedbl{}building blocks\textquotedbl{}
that are attached to each other in a prescribed way to form each of
the objects. The transplantation can be described in terms of these
building blocks. It expresses the restriction of an eigenfunction
to a building block of the first object as a linear combination of
the restrictions of an eigenfunction to building blocks of the second
one. The transplantation can be therefore described by a matrix whose
dimension equals to the number of the building blocks. Furthermore,
this matrix is independent of the spectral parameter $k$.

It is important to note that isospectrality does not necessarily imply transplantability. Namely, there are examples of isospectral objects for which there is no known transplantation. Such an example appears in section \ref{sec:lack_of_transplantability}.

\subsection{The scattering matrix of a quantum graph}

To convert a compact graph $\Gamma$ into a scattering system, one
can connect its vertices (all or a subset) by leads which extend to
infinity. We will denote by $\tg$ the extended quantum graph which
consists of the original graph $\Gamma$ and the external leads. Given
$\Gamma$, the additional information that one needs in order to describe
its extension $\tg$ consists of the set of vertices to which the
leads are connected and also, the modified vertex conditions at these
vertices. The vertices which are not connected to leads in $\tg$
have the same vertex conditions as they had in $\Gamma$. It is therefore
clear that there is more than one possible way to turn a compact graph
$\Gamma$ into a scattering system.

The graph $\tg$ has a continuous spectrum and we denote by $\Phi_{\tg}\left(k\right)$
the space of all generalized eigenfunctions of $\tg$ with eigenvalue
$k^{2}$: they are called generalized since they do not necessarily
have a bounded $L^{2}$-norm. Let $f\in\Phi_{\tg}\left(k\right)$
and let $\leads$ be the set of leads connected to $\Gamma$. Then
the restriction of $f$ to the lead $l\in\leads$ can be written in
the form

\begin{equation}
f_{l}\left(x_{l}\right)=a_{l}^{\, in}\exp\left(-ikx_{l}\right)+a_{l}^{\, out}\exp\left(ikx_{l}\right).\label{eq:function_on_leads}\end{equation}
Collecting all the variables $\left\{ a_{l}^{\, in}\right\} _{l\in\leads}$
and $\left\{ a_{l}^{\, out}\right\} _{l\in\leads}$ into the vectors
$\vec{a}^{\, in}$ and $\vec{a}^{\, out}$, we introduce the shorthand
notation\[
\left.f\right|_{\leads}=\vec{a}^{\, in}\exp\left(-ikx\right)+\vec{a}^{\, out}\exp\left(ikx\right).\]
 Using the vertex conditions on all the vertices of the graph, we
get equations whose variables contain $\left\{ a_{l}^{\, in}\right\} _{l\in\leads}$
and $\left\{ a_{l}^{\, out}\right\} _{l\in\leads}$. Simple algebraic
manipulations allow to find a relation of the following type\begin{equation}
\vec{a}^{\, out}=S_{\tg}\left(k\right)\vec{a}^{\, in},\label{eq:basic_scattering_relation}\end{equation}
where $S_{\tg}\left(k\right)$ is a square matrix of dimension $\left|\leads\right|$
and it is unitary for every $k\in\rr$. This is the scattering matrix
of the graph $\tg$. The existence and uniqueness of $S_{\tg}\left(k\right)$
for every value of $k$ and its unitarity on the real axis are proved
in \cite{BBS10}.

Once $S_{\tg}\left(k\right)$ is given, we may consider a graph $\tg_{0}$
which consists of a single vertex, connected to the same number of
leads as $\tg$ and with vertex conditions given by $S_{\tg}\left(k\right)$.
The graphs $\tg_{0}$ and $\tg$ share the same spectral properties
and their eigenfunctions coincide on the leads. Some of the information
about the graph's complex structure is contained within the function
$S_{\tg}\left(k\right)$. In particular, $S_{\tg}\left(k\right)$
contains the spectral information of $\Gamma$ as is presented by
the {}``exterior-interior\textquotedbl{} duality for graphs \cite{KS03}.
This means that the eigenvalues of $\Gamma$ can be identified as
the solutions of the equation \begin{equation}
\det(I-S_{\tg}\left(k\right))=0,\label{eq:secscat}\end{equation}
where $\tg$ is any extension of $\Gamma$ , with Neumann conditions
at the vertices attached to the leads.

All of the above makes one wonder about the data that is contained
in the scattering matrix and its comparison to the spectral
information. This motivates the following definitions
\footnote{Since there are various definitions in the literature, we
specify what isopolar, isophasal and isoscattering mean in our
setting.}
\begin{defn}
Let $\tg_{1}$ and $\tg_{2}$ be two quantum graphs. They are called
$isopolar$ if their scattering matrices
$S_{\tg_{1}}\left(k\right),S_{\tg_{2}}\left(k\right)$ share the same
poles on the complex plane.
\end{defn}
\begin{defn}
Let $\tg_{1}$ and $\tg_{2}$ be two quantum graphs. They are called
$isophasal$ if they satisfy the following condition
\begin{equation}\label{isophasal}
\forall k\in
\mathbb{R}\quad\frac{1}{i}\log(\mathrm{det}S_{\tg_{1}}\left(k\right))=\frac{1}{i}\log(\mathrm{det}S_{\tg_{2}}\left(k\right)).
\end{equation}
\end{defn}
\noindent We will use the term $isoscattering$ to refer to either
$isopolar$ or $isophasal$ quantum graphs. The present paper treats
the construction of isoscattering graphs and the relation of
isospectrality to isoscattering.

\section{A basic example \label{sec:A-basic-example}}

\begin{figure}[h]
\includegraphics[scale=0.2]{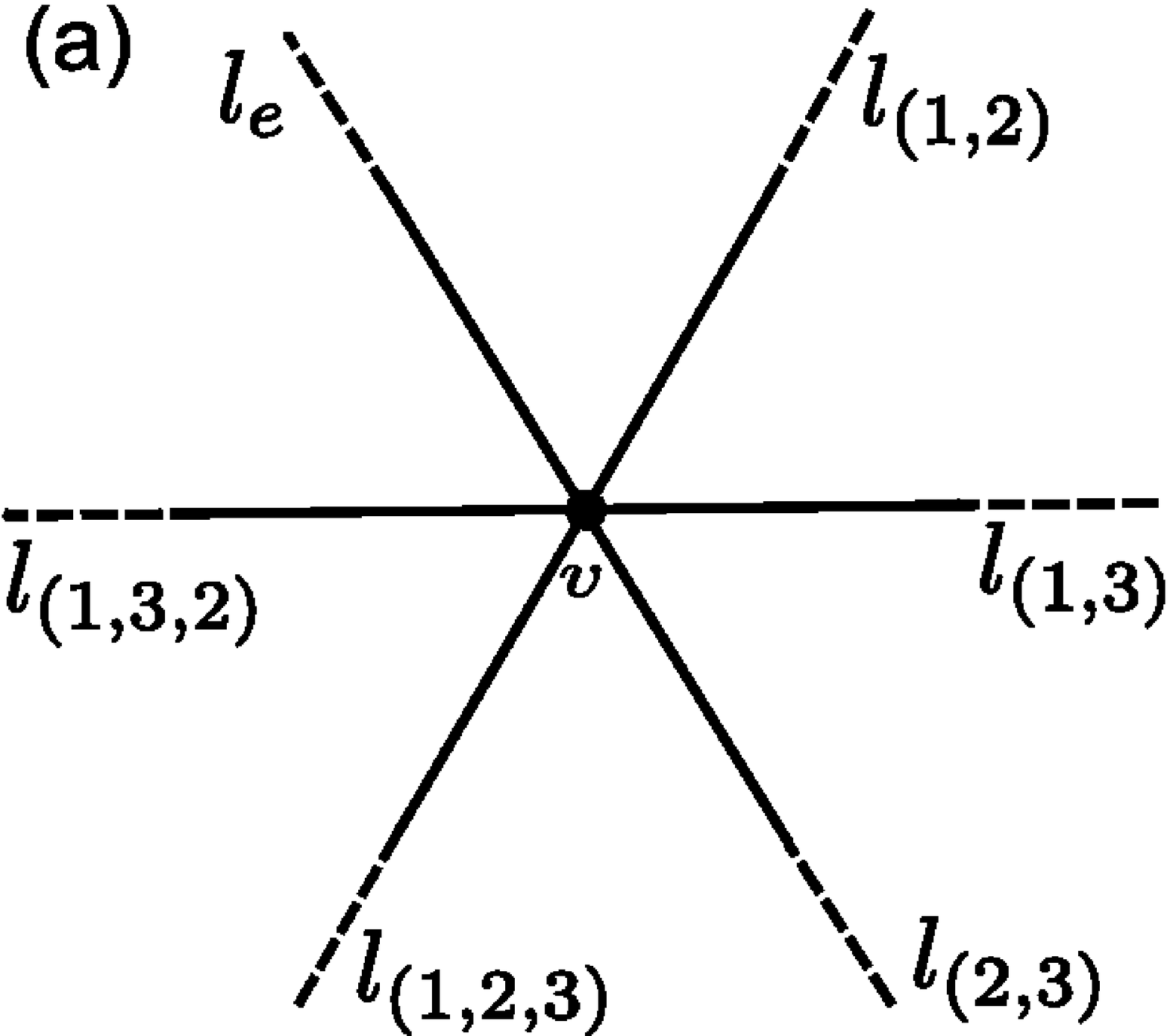}\qquad{}\qquad{}\qquad{}\qquad{}\qquad{}\qquad{}\qquad{}\includegraphics[scale=0.2]{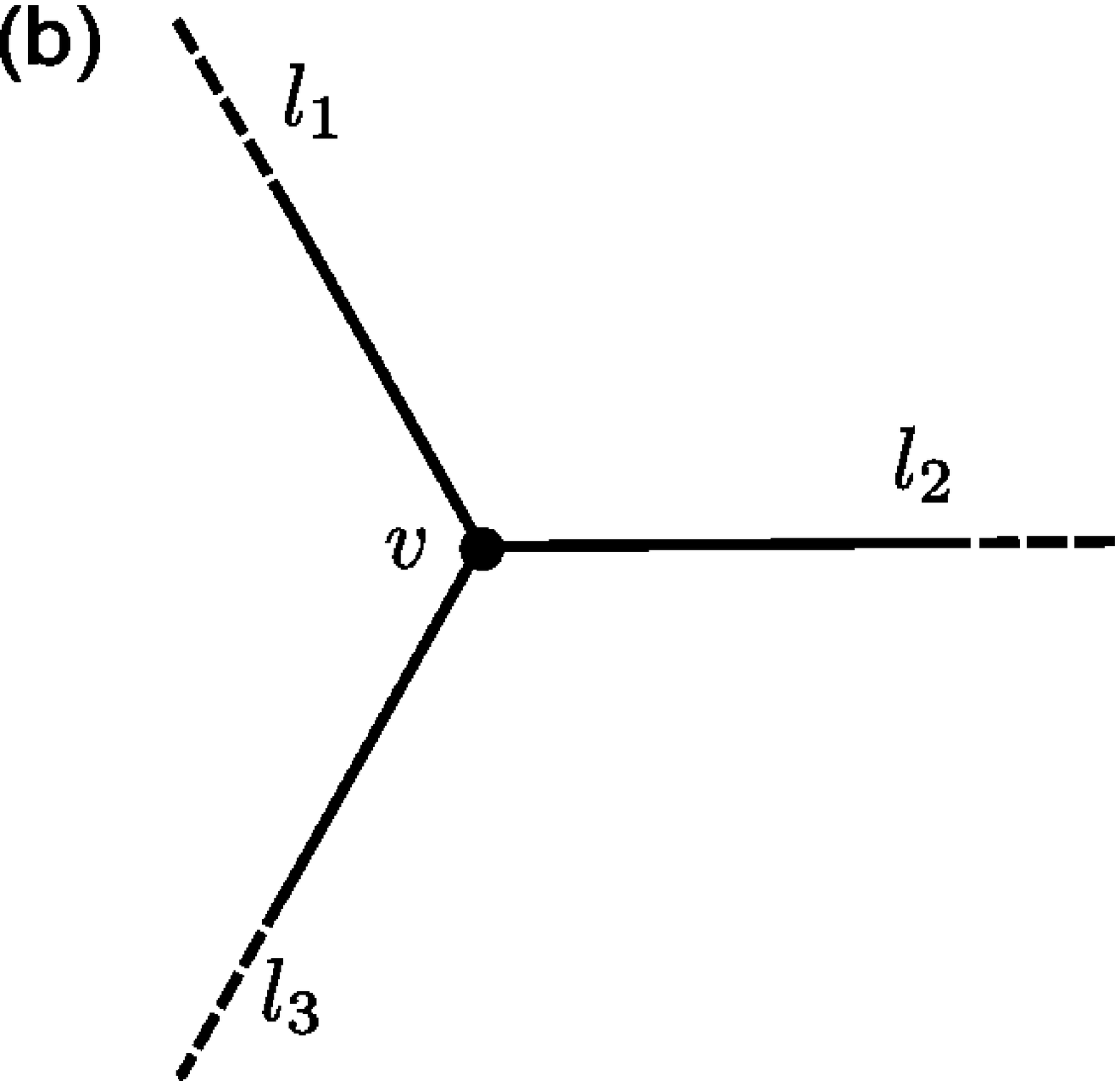}\caption{(a) The graph $\tilde{\Gamma}$ with leads labeled by the elements
of $G=S_{3}$ (b) The graph $\nicefrac{\tilde{\Gamma}}{\mathbf{1}_{H}}$.}
\label{Flo:rys}
\end{figure}

In this section we explain using a simple example the idea of the
quotient graph construction. We then use this example throughout the
paper to illustrate its main results. Following the discussion after
equation (\ref{eq:basic_scattering_relation}) we can restrict our
attention to graphs with a single vertex attached to leads. Let us
consider the quantum graph $\tilde{\Gamma}$ that consists of six
semi-infinite leads which are connected to a single vertex $v$. The
vertex conditions at $v$ are Neumann conditions (see figure \ref{Flo:rys}(a)).

\noindent We label $\Gamma$'s leads by the elements of the group
$G=S_{3}$, and parameterize each lead $l_{g}$ by a coordinate
$x_{g}\in[0,\infty[$ such that $x_{g}(v)=0$. The permutation group
$G$ acts on the graph in the following way:\begin{eqnarray} \forall
g,h & \in & G,\quad\forall x_{h}\in l_{h}:\quad gx_{h}=x_{gh}\in
l_{gh}.\label{eq:action}\end{eqnarray}
 The action of $g\in G$ on $f\in\Phi_{\tilde{\Gamma}}(k)$ gives
the function $gf\in\Phi_{\tilde{\Gamma}}(k)$ defined by:

\begin{equation}
(gf)(x)=f(g^{-1}x).\end{equation}
\noindent Let us consider the subgroup $H=\{e,(1,2)\}$ of $G$.
The action of $H$ on the leads of $\tilde{\Gamma}$ gives the following
orbits: \[
\mathcal{O}_{e}=\{l_{e},l_{(1,2)}\},\quad\mathcal{O}_{(1,3)}=\{l_{(1,3)},l_{(1,3,2)}\},\quad\mathcal{O}_{(1,2,3)}=\{l_{(1,2,3)},l_{(2,3)}\}.\]
Let us restrict our attention to those functions $f\in\Phi_{\tilde{\Gamma}}(k)$
which transform under the action of $H$ according to the trivial
representation $\mathbf{1}_{H}$ of $H$. This means that:

\[
f|_{l_{e}}=f|_{l_{(1,2)}},\quad f|_{l_{(1,3)}}=f|_{l_{(1,3,2)}},\quad f|_{l_{(1,2,3)}}=f|_{l_{(2,3)}}.\]
\noindent Therefore, it is enough to know the values of the function
$f$ on the three leads $l_{e},l_{(1,3)}$ and $l_{(1,2,3)}$, being
representatives of each orbit, to deduce its values on the whole of
$\tilde{\Gamma}.$ We wish to encode the information about such a
function and may do so by constructing the so-called quotient graph.
The quotient, which we denote by $\nicefrac{\tilde{\Gamma}}{\mathbf{1}_{H}}$
has three leads, each is a representative of an orbit of the $H$-action
(figure \ref{Flo:rys}(b)). The encoding is described by the map $\phi$
that acts on functions which transforms under the action of $H$ according
to representation $\mathbf{1}_{H}$. It takes the values of such $f\in\Phi_{\tilde{\Gamma}}(k)$
restricted to the leads $l_{e}$, $l_{(1,2)}$, $l_{(1,2,3)}$ and
assigns them to the function $\phi f$ on the leads $l_{1}$, $l_{2}$,
$l_{3}$ of $\nicefrac{\tilde{\Gamma}}{\mathbf{1}_{H}}$ . By the
construction we see that $\phi f$ satisfies the following vertex
conditions\[
(\phi f)|_{l_{1}}(v)=(\phi f)|_{l_{2}}(v)=(\phi f)|_{l_{3}}(v)\]

\[
\sum_{i=1}^{3}(\phi f)|_{l_{i}}^{\prime}(v)=f|_{l_{e}}^{\prime}(v)+f|_{l_{(1,3)}}^{\prime}(v)+f|_{l_{(1,2,3)}}^{\prime}(v)=\frac{1}{2}\sum_{g\in G}f|_{l_{g}}^{\prime}(v)=0.\]
We therefore equip $\nicefrac{\tilde{\Gamma}}{\mathbf{1}_{H}}$ with
Neumann condition at its vertex. In general the vertex conditions
at each vertex of a quotient graph $\nicefrac{\tilde{\Gamma}}{R}$
are determined by the vertex conditions of the corresponding vertices
in $\tilde{\Gamma}$ combined with the group action and the representation
$R$. The exact formula for the general case is given in \cite{PB09}.

\begin{figure}[h]
\includegraphics[scale=0.4]{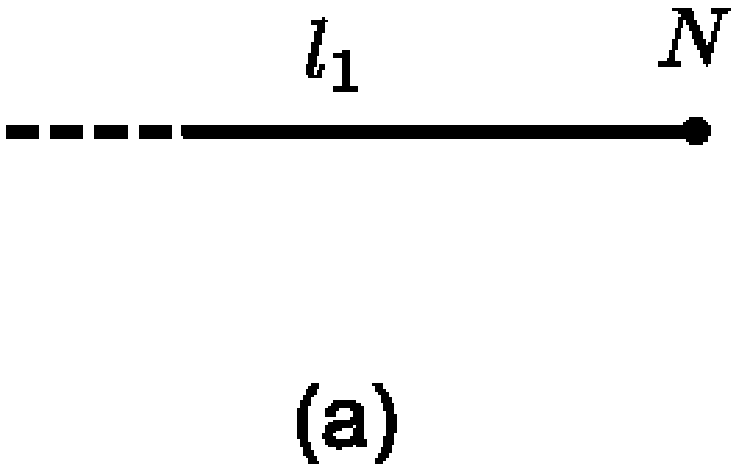}\qquad{}\includegraphics[scale=0.4]{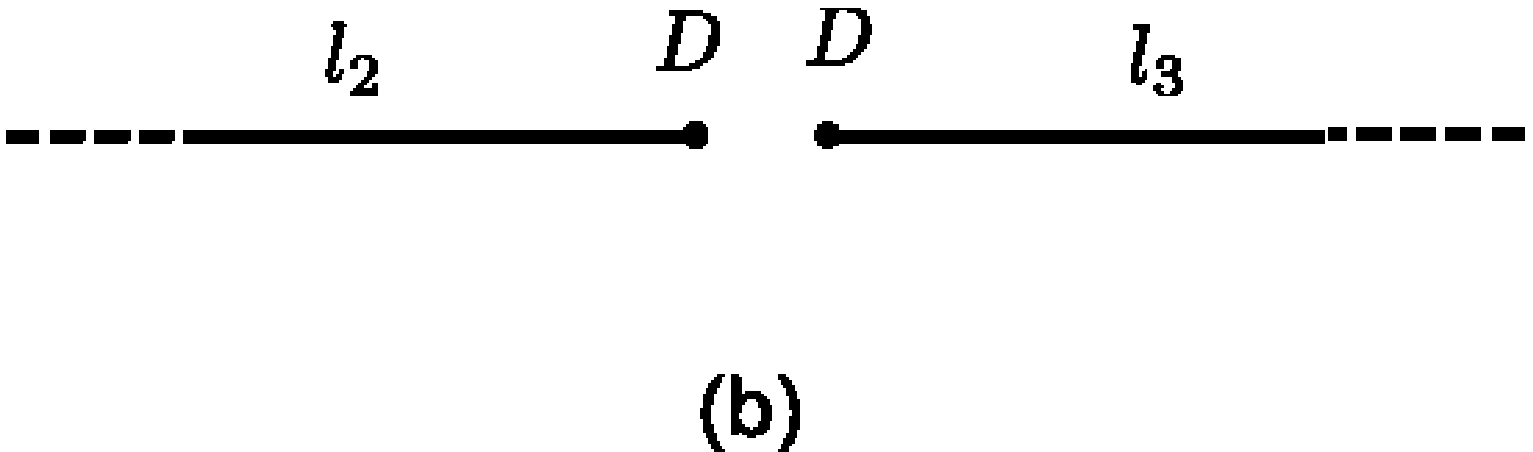}

\caption{(a) The graph $\nicefrac{\tilde{\Gamma}}{\mathbf{1}_{G}}$ (b) The
graph $\nicefrac{\tilde{\Gamma}}{R_{2d}}$. In both cases $N$ stands
for Neumann vertex conditions and $D$ for Dirichlet vertex conditions}
\label{Flo:qR1R2}
\end{figure}

We now construct another quotient graph which will be shown to be
isoscattering to $\nicefrac{\tilde{\Gamma}}{\mathbf{1}_{H}}$. The
quotient will be calculated with respect to the representation $R=\mathbf{1}_{G}\oplus R_{2d}$,
where $\mathbf{1}_{G}$ is the trivial representation of $G$ and
$R_{2d}$ is the two dimensional irreducible representation of $G$.
It was shown in \cite{BPB09} that $\nicefrac{\tilde{\Gamma}}{R}=\nicefrac{\tilde{\Gamma}}{\mathbf{1}_{G}}\cup\nicefrac{\tilde{\Gamma}}{R_{2d}}$
and therefore we may construct each of the quotients on the right
hand side separately. The quotient $\nicefrac{\tilde{\Gamma}}{\mathbf{1}_{G}}$,
can be inferred using similar reasoning as in the case of $\nicefrac{\tilde{\Gamma}}{\mathbf{1}_{H}}$,
and it turns out to be a single lead with Neumann vertex condition
(figure \ref{Flo:qR1R2}(a)). In order to construct $\nicefrac{\tilde{\Gamma}}{R_{2d}}$
we choose a basis in $V^{R_{2d}}\cong\mathbb{C}^{2}$ and write the
corresponding matrix representation $\rho^{R_{2d}}:G\rightarrow Gl(2,\mathbb{C})$
of $R_{2d}$. Such a choice gives for example:

\scriptsize

\begin{eqnarray}
\rho^{R_{2d}}(e) & = & \left(\begin{array}{cc}
1 & 0\\
0 & 1\end{array}\right),\quad\rho^{R_{2d}}((1,2))=\left(\begin{array}{cc}
-1 & 1\\
0 & 1\end{array}\right),\quad\rho^{R_{2d}}((1,3))=\left(\begin{array}{cc}
0 & -1\\
-1 & 0\end{array}\right),\nonumber \\
\rho^{R_{2d}}((2,3)) & = & \left(\begin{array}{cc}
1 & 0\\
1 & -1\end{array}\right),\quad\rho^{R_{2d}}((1,2,3))=\left(\begin{array}{cc}
0 & -1\\
1 & -1\end{array}\right),\quad\rho^{R_{2d}}((1,3,2))=\left(\begin{array}{cc}
-1 & 1\\
-1 & 0\end{array}\right).\label{eq:mat}\end{eqnarray}

\normalsize

\medskip{}

\noindent The representation $R_{2d}$ is two dimensional, which
means we need to consider two linearly independent functions $f_{1},f_{2}\in\Phi_{\tilde{\Gamma}}(k)$
that transform under the action of $G$ according to the representation
$R_{2d}$, and in particular

\begin{eqnarray}
\forall g & \in & G\quad[f_{1}|_{l_{g}},f_{2}|_{l_{g}}]^{T}=(\rho^{R_{2d}}(g))[f_{1}|_{l_{e}},f_{2}|_{l_{e}}]^{T}.\label{eq:rep2d}\end{eqnarray}
The restriction of $f_{1}$ and $f_{2}$ to $l_{e}$ together with
property (\ref{eq:rep2d}) can be used to determine the values of
$f_{1},f_{2}$ on the whole of $\tilde{\Gamma}$. The encoding process
therefore requires two leads, which is the number of leads in the
quotient graph $\nicefrac{\tilde{\Gamma}}{R_{2d}}$. It is only left
to conclude the vertex conditions of this graph. Plugging $g=(1,3)$
and $g=(2,3)$ in (\ref{eq:rep2d}) we obtain: \begin{eqnarray}
[f_{1}|_{l_{(1,3)}},f_{2}|_{l_{(1,3)}}]^{T} & = & [-f_{2}|_{l_{e}},-f_{1}|_{l_{e}}]^{T}\label{eq:verte}\\
{}[f_{1}|_{l_{(2,3)}},f_{2}|_{l_{(2,3)}}]^{T} & = & [f_{1}|_{l_{e}},f_{1}|_{l_{e}}-f_{2}|_{l_{e}}]^{T}\label{eq:verte1}\end{eqnarray}
Restricting (\ref{eq:verte}), (\ref{eq:verte1}) to the vertex $v$
which obeys Neumann conditions we see that their left hand sides are
equal to each other. Carrying this to the right hand sides we get:
\begin{eqnarray}
-f_{2}|_{l_{e}}(v) & = & f_{1}|_{l_{e}}(v)\\
-f_{1}|_{l_{e}}(v) & = &
f_{1}|_{l_{e}}-f_{2}|_{l_{e}}(v)\end{eqnarray} These two relations
give $f_{1}(v)=0$ and $f_{2}(v)=0$. One can check that using the
other group elements we do not add any more linearly independent
conditions on the values or on the derivatives of $f_{1}$ and
$f_{2}$. We therefore conclude that
$\nicefrac{\tilde{\Gamma}}{R_{2d}}$ is the union of two leads with
Dirichlet vertex conditions as is shown on figure
\ref{Flo:qR1R2}(b).

\section{Isospectrality and scattering matrices\label{sec:isospectrality_and_scattering}}

\subsection{The scattering matrix of the quotient graph\label{sec:scat_of_quotient}}

In the following section we apply the isospectral construction presented
in \cite{BPB09,PB09} to quantum graphs with leads. This allows to
investigate the relation between scattering matrices of isospectral
graphs. Let $\Gamma$ be a quantum graph and $\tg$ some extension
of $\Gamma$ to a scattering system. Let $G$ be a symmetry group
of $\tg$ and let $R$ be a matrix representation of $G$ with some
carrier space $V^{R}$. We may then apply the isospectral theory to
have the analogue of (\ref{eq:quotietgreigen}) for graphs with leads:
\begin{equation}
\Psi:\Phi_{\nicefrac{\tg}{R}}\left(k\right)\overset{\cong}{\longrightarrow}\mathrm{Hom}_{\mathbb{C}G}\left(V^{R},\Phi_{\tg}\left(k\right)\right).\label{eq:quotient_graph_with_leads_eigenspace}\end{equation}
Choosing some $v\in V^{R}$, we define

\begin{eqnarray*}
\Psi^{\left(v\right)}:\Phi_{\nicefrac{\tg}{R}}\left(k\right)\rightarrow\Phi_{\tg}\left(k\right)\\
\Psi^{\left(v\right)}\left(f\right):=\Psi\left(f\right)\left(v\right),\end{eqnarray*}
and denote the image of $\Psi^{(v)}$ by

\[
\Phi_{\tg}^{\left(v\right)}\left(k\right):=\left\{ \left.\Psi\left(f\right)\left(v\right)\right|\, f\in\Phi_{\nicefrac{\tg}{R}}\left(k\right)\right\} .\]
It is easy to see that $\Phi_{\tg}^{(v)}(k)$ is a vector space. One
can also show that if $\mathrm{Hom}_{\mathbb{C}G}\left(V^{R},\Phi_{\tg}\left(k\right)\right)$
is non-trivial and $R$ is an irreducible representation, then the
subset $\mathcal{F}\subset\Phi_{\nicefrac{\tg}{R}}\left(k\right)$
is linearly independent if and only if $\left\{ \Psi^{(v)}\left(f\right)\right\} _{f\in\mathcal{F}}$
is linearly independent. We therefore restrict ourselves for the moment
to the case of irreducible representations, for which we get

\begin{equation}
\Psi^{\left(v\right)}:\Phi_{\nicefrac{\tg}{R}}\left(k\right)\overset{\cong}{\longrightarrow}\Phi_{\tg}^{\left(v\right)}\left(k\right).\label{eq:phivcomponent}\end{equation}
Having in mind that this isospectral theory is motivated by an
encoding scheme, we give such an interpretation to
(\ref{eq:phivcomponent}). Each eigenfunction of the quotient
$\nicefrac{\tg}{R}$ encodes information on some eigenfunction of
$\tg$ which transforms just like $v$ transforms, under the group
action, as it is described by the matrix representations $R$.

In order to relate this construction to the corresponding scattering
matrices, we need to rephrase the equations above in terms of the
spaces of ingoing and outgoing amplitudes on the leads rather then
the eigenspaces. In order to do so, we introduce the notation $L_{\tg}$
for the vector space in which $\vec{a}^{\, in}$ lies. It is easy
to see that given $f\in\Phi_{\tg}\left(k\right)$, the corresponding
vector $\vec{a}^{\, in}\in L_{\tg}$ can be uniquely determined by
restricting $f$ to the leads and reading the coefficients of the
exponents $\exp\left(-ikx_{l}\right)$. In addition, it was shown
in \cite{BBS10} that the opposite also holds, i.e., $\vec{a}^{\, in}\in L_{\tg}$
uniquely determines the corresponding $f\in\Phi_{\tg}\left(k\right)$%
\footnote{This is true under some generality conditions which are stated in
\cite{BBS10}. %
}. We therefore have a natural ($k$-dependent) isomorphism between
those spaces \begin{eqnarray}
L_{\tg}\cong\Phi_{\tg}\left(k\right).\label{eq:leads_cong_phi}\end{eqnarray}
One should note that there is a well defined action of $G$ on $L_{\tg}$,
as it induces the action of $G$ on the leads of $\tg$. The isomorphism
in (\ref{eq:leads_cong_phi}) is actually an intertwiner, and as such,
we can rewrite all the equalities in this section in terms of $L_{\tg}$
and $L_{\nicefrac{\tg}{R}}$. In particular we get the following analogue
of (\ref{eq:phivcomponent})

\begin{equation}
\Upsilon^{\left(v\right)}:L_{\nicefrac{\tg}{R}}\overset{\cong}{\longrightarrow}L_{\tg}^{\left(v\right)},\label{eq:L_v_component}\end{equation}
where $\Upsilon^{\left(v\right)}$ is the restriction of $\Psi^{\left(v\right)}$
on the spaces $L_{\nicefrac{\tg}{R}}$ and $L_{\tg}$, and $L_{\tg}^{\left(v\right)}$
is the pre-image of $\Psi^{(v)}$ under the isomorphism in (\ref{eq:leads_cong_phi}).

An important property of the encoding (\ref{eq:phivcomponent}) (also
(\ref{eq:quotient_graph_with_leads_eigenspace})) is that it does
not mix between the ingoing amplitudes $\vec{a}^{\, in}$ and the
outgoing ones, $\vec{a}^{\, out}$. This results from the nature of
the encoding and means that the ingoing and the outgoing amplitudes
are encoded in exactly the same manner. Namely, for some $f\in\Phi_{\nicefrac{\tg}{R}}\left(k\right)$
whose values on the leads are

\[
\left.f\right|_{\leads}=\vec{a}^{\, in}\exp\left(-ikx\right)+\vec{a}^{\, out}\exp\left(ikx\right),\]
 we have \begin{equation}
\left.\left(\Psi^{\left(v\right)}f\right)\right|_{\leads}=\Upsilon^{\left(v\right)}\left(\vec{a}^{\, in}\right)\exp\left(-ikx\right)+\Upsilon^{\left(v\right)}\left(\vec{a}^{\, out}\right)\exp\left(ikx\right).\label{eq:Psi_and_Upsilon}\end{equation}

Note that in (\ref{eq:Psi_and_Upsilon}) we consider the spaces $L_{\tg}$,
$L_{\nicefrac{\tg}{R}}$ as spaces which contain not only the ingoing
amplitudes vectors $\vec{a}{}^{\, in}$, but also the outgoing ones,
$\vec{a}{}^{\, out}$. We now have all that is required to express
$S_{\nicefrac{\tg}{R}}$ in terms of $S_{\tg}$. Let $\vec{a}_{\nicefrac{\tg}{R}}^{\, in}\in\lgr$.
There exists a corresponding function $f\in\Phi_{\nicefrac{\tg}{R}}\left(k\right)$,
whose restriction to the leads is given by \[
\left.f\right|_{\leads}=\vec{a}_{\nicefrac{\tg}{R}}^{\, in}\exp\left(-ikx\right)+S_{\nicefrac{\tg}{R}}\vec{a}_{\nicefrac{\tg}{R}}^{\, in}\exp\left(ikx\right).\]
Applying $\Psi^{\left(v\right)}$ (with an arbitrary $v\in V^{R}$)
to $f$ and restricting again to the leads (this time the leads of
$\tg$) we get\begin{equation}
\left.\left(\Psi^{\left(v\right)}f\right)\right|_{\mathcal{L}}=\vec{a}_{\tg}^{\, in}\exp\left(-ikx\right)+S_{\tg}\vec{a}_{\tg}^{\, in}\exp\left(ikx\right),\label{eq:Psi_of_f}\end{equation}
where $\vec{a}_{\tg}^{\, in}$ is an appropriately chosen vector in
$L_{\tg}$. Comparing (\ref{eq:Psi_of_f}) to (\ref{eq:Psi_and_Upsilon})
gives\begin{eqnarray*}
\vec{a}_{\tg}^{\, in} & = & \Upsilon^{\left(v\right)}\left(\vec{a}_{\nicefrac{\tg}{R}}^{\, in}\right)\\
S_{\tg}\vec{a}_{\tg}^{\, in} & = & \Upsilon^{\left(v\right)}\left(S_{\nicefrac{\tg}{R}}\vec{a}_{\nicefrac{\tg}{R}}^{\, in}\right).\end{eqnarray*}
From the above we arrive to \[
S_{\nicefrac{\tg}{R}}\vec{a}_{\nicefrac{\tg}{R}}^{\, in}=\left(\Upsilon^{\left(v\right)}\right)^{-1}S_{\tg}\Upsilon^{\left(v\right)}\left(\vec{a}_{\nicefrac{\tg}{R}}^{\, in}\right),\]
which holds for every $\vec{a}_{\nicefrac{\tg}{R}}^{\, in}\in L_{\nicefrac{\tilde{\Gamma}}{R}}$
and therefore\begin{equation}
S_{\nicefrac{\tg}{R}}=\left(\Upsilon^{\left(v\right)}\right)^{-1}S_{\tg}\Upsilon^{\left(v\right)}.\label{eq:scattering_of_quotient}\end{equation}

We can generalize (\ref{eq:scattering_of_quotient}) to any representation
$R=\oplus_{i}n_{i}R_{i}$, where $\left\{ R_{i}\right\} $ are the
irreducible representations of $G$. We choose a vector $v_{i}$ from
each of the abstract carrier spaces $\left\{ V^{R_{i}}\right\} $
of the representations $\left\{ R_{i}\right\} $. The scattering matrix
$S_{\nicefrac{\tg}{R}}$ is then composed of the sub-matrices $\left(\Upsilon^{(v_{i})}\right)^{-1}S_{\tg}\Upsilon^{(v_{i})}$,
each appearing $n_{i}$ times.

\subsection{Back to the example}

We now implement (\ref{eq:scattering_of_quotient}) in order to calculate
the scattering matrices of the quotient graphs $\nicefrac{\tilde{\Gamma}}{\mathbf{1}_{H}}$,
$\nicefrac{\tilde{\Gamma}}{R}$ presented in section \ref{sec:A-basic-example}.
The scattering matrix of a star graph with Neumann vertex conditions
is given by \begin{eqnarray}
S_{ij} & = & \frac{2}{d}-\delta_{ij},\label{eq:starscat}\end{eqnarray}
where $d$ is the valency of graph's vertex. Applying this formula
to the graph $\tilde{\Gamma}$ (where $d=6$) we get the scattering
matrix $(S_{\tilde{\Gamma}})_{ij}=\frac{1}{3}-\delta_{ij}$. The space
of ingoing\textbackslash{}outgoing amplitudes in this case is $L_{\tilde{\Gamma}}=\mathbb{C}^{6}$.
We denote the standard basis elements of $\mathbb{C}^{6}$ by the
elements of $G=S_{3}$ and infer using equation (\ref{eq:action})
and figure \ref{Flo:rys}(a) that the subgroup $H=\{e,(1,2)\}$ is
acting on $L_{\tilde{\Gamma}}$ by:

\scriptsize

\begin{eqnarray}
e & \rightarrow & \mathbf{1}_{L_{\tilde{\Gamma}}},\quad(1,2)\rightarrow\left(\begin{array}{cccccc}
0 & 1 & 0 & 0 & 0 & 0\\
1 & 0 & 0 & 0 & 0 & 0\\
0 & 0 & 0 & 0 & 0 & 1\\
0 & 0 & 0 & 0 & 1 & 0\\
0 & 0 & 0 & 1 & 0 & 0\\
0 & 0 & 1 & 0 & 0 & 0\end{array}\right),\label{eq:Hact}\end{eqnarray}

\normalsize In order to compute $L_{\nicefrac{\tilde{\Gamma}}{\mathbf{1}_{H}}}$
we need to find a subspace of linearly independent vectors in $L_{\tilde{\Gamma}}$
that transform according to the trivial representation under the action
of $H$. In other words, we are looking for eigenvectors of the matrix
assigned to $(1,2)$ by (\ref{eq:Hact}) with eigenvalue one. This
subspace is:

\[
L_{\tilde{\Gamma}}^{(v)}\mathrm{=Span}([1,1,0,0,0,0]^{T},[0,0,1,0,0,1]^{T},[0,0,0,1,1,0]^{T}).\]
This is actually the image of $\Upsilon^{(v)}$ acting on $L_{\nicefrac{\tilde{\Gamma}}{\mathbf{1}_{H}}}$
and therefore a possible basis for $L_{\nicefrac{\tilde{\Gamma}}{\mathbf{1}_{H}}}$
is given by the action of encoding map $(\Upsilon^{(v)})^{-1}$ on
the vectors above:

\begin{eqnarray*}
(\Upsilon^{(v)})^{-1}[1,1,0,0,0,0]^{T} & = & [1,0,0]^{T},\\
(\Upsilon^{(v)})^{-1}[0,0,1,0,0,1]^{T} & = & [0,1,0]^{T},\\
(\Upsilon^{(v)})^{-1}[0,0,0,1,1,0]^{T} & = & [0,0,1]^{T},\end{eqnarray*}
where the encoding is just a projection on the first, third and fifth
coordinates of $L_{\tilde{\Gamma}}$ as was explained in section \ref{sec:A-basic-example}.
Making use of (\ref{eq:scattering_of_quotient}) we get: \begin{eqnarray}
S_{\nicefrac{\tilde{\Gamma}}{\mathbf{1}_{H}}}(k) & = & \left(\Upsilon^{(v)}\right)^{-1}S_{\tg}(k)\Upsilon^{(v)}=\frac{1}{3}\left(\begin{array}{ccc}
-1 & 2 & 2\\
2 & -1 & 2\\
2 & 2 & -1\end{array}\right).\label{eq:scatmat1H}\end{eqnarray}
As expected, the above corresponds the general formula for the scattering
matrix of a star graph with Neumann conditions (\ref{eq:starscat}).

We know that for $\nicefrac{\tilde{\Gamma}}{R}$ the scattering matrix
is the direct sum $S_{\nicefrac{\tilde{\Gamma}}{\mathbf{1}_{G}}}\oplus S_{\nicefrac{\tilde{\Gamma}}{R_{2d}}}$.
The first element in the sum is $S_{\nicefrac{\tilde{\Gamma}}{\mathbf{1}_{G}}}=1$
(by similar reasoning as for $\nicefrac{\tilde{\Gamma}}{\mathbf{1}_{H}}$).
In order to compute $L_{\nicefrac{\tilde{\Gamma}}{R_{2d}}}$ we need
to choose any vector $v$ from the abstract carrier space $V^{R_{2d}}$
and to find a subspace of linearly independent vectors in $L_{\tilde{\Gamma}}$
that transform under the action of $G$ as $v$ under the action of
$R_{2d}$ (given by the matrices (\ref{eq:mat})). The vector that
we choose is the first basis vector for which $R_{2d}$ is represented
by the matrices (\ref{eq:mat}). It can be easily checked that there
are exactly two linearly independent vectors in $L_{\tilde{\Gamma}}$
that fulfill this conditions:

\[
L_{\tilde{\Gamma}}^{(v)}=\mathrm{Span}([1,-1,0,1,1,-1]^{T},[0,1,-1,0,-1,1]^{T}).\]
The action of the encoding map $(\Upsilon^{(v)})^{-1}$ on these vectors
gives:

\begin{eqnarray*}
(\Upsilon^{(v)})^{-1}[1,-1,0,1,1,-1]^{T} & = & [1,0]^{T},\\
(\Upsilon^{(v)})^{-1}[0,1,-1,0,-1,1]^{T} & = & [0,1]^{T}.\end{eqnarray*}
Finally

\[
S_{\nicefrac{\tilde{\Gamma}}{R_{2d}}}(k)=\left(\Upsilon^{(v)}\right)^{-1}S_{\tg}(k)\Upsilon^{(v)}=\left(\begin{array}{cc}
-1 & 0\\
0 & -1\end{array}\right),\]
 hence

\begin{eqnarray}
S_{\nicefrac{\tilde{\Gamma}}{R}}(k) & = & \left(\begin{array}{ccc}
1 & 0 & 0\\
0 & -1 & 0\\
0 & 0 & -1\end{array}\right).\label{eq:scatmathR2d}\end{eqnarray}

In the example above we have calculated the scattering matrices of
the quotient graphs $\nicefrac{\tilde{\Gamma}}{\mathbf{1}_{H}}$,
$\nicefrac{\tilde{\Gamma}}{\mathbf{1}_{G}}$, $\nicefrac{\tilde{\Gamma}}{R_{2d}}$
using the formula (\ref{eq:scattering_of_quotient}). Furthermore,
this is actually a reconstruction of the graphs themselves, as we
now explain. Each of the graphs $\nicefrac{\tilde{\Gamma}}{\mathbf{1}_{H}}$,
$\nicefrac{\tilde{\Gamma}}{\mathbf{1}_{G}}$, $\nicefrac{\tilde{\Gamma}}{R_{2d}}$
consists of a single vertex attached to leads and therefore its scattering
matrix dictates the vertex conditions and gives a complete description
of the graph. This can be generalized to form an alternative description
for the quotient graph construction in \cite{BPB09,PB09}. A fundamental
element of the quotient construction is the ability to express the
vertex conditions of the quotient in terms of the vertex conditions
of the original graph and the group action. The vertex conditions
in \cite{BPB09,PB09} are described by stating linear sets of equations
on the values and the derivatives of the function at each vertex.
The quotient construction relates the quotient equation sets to those
of the original graph. However, as was explained in section \ref{sec:quantum_graphs},
vertex conditions can be also represented by unitary scattering matrices
at the vertices. Using this approach one can construct a quotient
graph by relating the scattering matrices at the vertices of the quotient
to the ones at the vertices of the original graph, using formula (\ref{eq:scattering_of_quotient})
and its generalization. The method of determining the exact structure
of the quotient graph, i.e., its edges, vertices and their connectivity
remains the same as in \cite{BPB09,PB09}. The difference in the current
approach is the ability to use the scattering description of the vertex
conditions in the construction process. This supplies us with more
graphs to which the construction can be applied, as there are vertex
conditions that can be described only using the scattering approach.
An example for such vertex conditions is given by the discrete Fourier
transform matrices\[
\sigma_{p,q}=\frac{1}{\sqrt{d}}\exp\left(2\pi i\,\frac{pq}{d}\right),\]
 where $d$ is the degree of the vertex (more details can be found
in \cite{GS06}).

\subsection{Isoscattering graphs\label{sec:isoscattering_graphs}}

Let $\Gamma_{1},\,\Gamma_{2}$ be isospectral graphs constructed by
the method given in \cite{BPB09,PB09}. Let $\Gamma$ be the graph
from which $\Gamma_{1},\,\Gamma_{2}$ are obtained as quotients. Namely,
there is a group $G$ and two subgroups $H_{1},H_{2}<G$ with corresponding
representations $R_{1},R_{2}$, such that $\Gamma_{1}=\nicefrac{\Gamma}{R_{1}},\,\Gamma_{2}=\nicefrac{\Gamma}{R_{2}}$
and $\textrm{Ind}_{H_{1}}^{G}R_{1}\cong\textrm{Ind}_{H_{2}}^{G}R_{2}$.
Let $\tilde{\Gamma}$ be a quantum graph which is obtained by attaching
leads to $\Gamma$ in a way that conserves both symmetries $H_{1},H_{2}$.
We may therefore construct the quotients $\tg_{1}=\nicefrac{\tilde{\Gamma}}{R_{1}},\,\tg_{2}=\nicefrac{\tilde{\Gamma}}{R_{2}}$
according to the method presented in \cite{BPB09,PB09} and obtain
the existence of the following transplantation \[
T:\Phi_{\tilde{\Gamma}_{1}}\left(k\right)\overset{\cong}{\longrightarrow}\Phi_{\tilde{\Gamma}_{2}}\left(k\right).\]

It was already mentioned in section \ref{sec:scat_of_quotient} that
the maps we consider can be restricted from the eigenspaces $\Phi_{\tg_{1}}$,
$\Phi_{\tg_{2}}$ to the leads' amplitudes spaces $L_{\tg_{1}}$,
$L_{\tg_{2}}$. Furthermore, they act the same on the ingoing and
on the outgoing amplitudes (recall (\ref{eq:Psi_and_Upsilon}) and
the explanation that precedes it). Exploiting this, we may apply the
transplantation on \[
\left.f\right|_{\leads}=\vec{a}_{\tg_{1}}^{\, in}\exp\left(-ikx\right)+S_{\tg_{1}}\vec{a}_{\tg_{1}}^{\, in}\exp\left(ikx\right),\]
 and get\begin{equation}
\left.\left(Tf\right)\right|_{\mathcal{L}}=\Pi\,\vec{a}_{\tg_{1}}^{\, in}\exp\left(-ikx\right)+\Pi\, S_{\tg_{1}}\vec{a}_{\tg_{1}}^{\, in}\exp\left(ikx\right),\label{eq:T_of_f}\end{equation}
where

\[
\Pi:L_{\tg_{1}}\overset{\cong}{\longrightarrow}L_{\tg_{2}}\left(k\right),\]
is the restriction of the transplantation to the spaces $L_{\tg_{1}},\, L_{\tg_{2}}$.

We can write (\ref{eq:T_of_f}) in the form\[
\left.\left(Tf\right)\right|_{\mathcal{L}}=\vec{a}_{\tg_{2}}^{\, in}\exp\left(-ikx\right)+S_{\tg_{2}}\vec{a}_{\tg_{2}}^{\, in}\exp\left(ikx\right),\]
where $\vec{a}_{\tg_{2}}^{\, in}$ is an appropriately chosen vector
in $L_{\tg_{2}}$. Comparing the two expressions for $\left.\left(Tf\right)\right|_{\mathcal{L}}$
we obtain\begin{eqnarray*}
\vec{a}_{\tg_{2}}^{\, in} & = & \Pi\,\vec{a}_{\tg_{1}}^{\, in}\\
S_{\tg_{2}}\vec{a}_{\tg_{2}}^{\, in} & = & \Pi\, S_{\tg_{1}}\vec{a}_{\tg_{1}}^{\, in},\end{eqnarray*}
which holds for every $\vec{a}_{\tg_{1}}^{\, in}\in L_{\tg}$ and
therefore\begin{equation}
S_{\tg_{1}}=\Pi^{-1}S_{\tg_{2}}\Pi.\label{eq:scattering_of_isospectral}\end{equation}

We conclude that the scattering matrices $S_{\tg_{1}}(k),\,
S_{\tg_{2}}(k)$ of the quotients
$\tg_{1}=\nicefrac{\tilde{\Gamma}}{R_{1}}$,
$\tg_{2}=\nicefrac{\tilde{\Gamma}}{R_{2}}$ are conjugate to each
other. Note that the conjugating matrix $\Pi$ does not depend on
$k$, as it inherits this property from the transplantation. In
particular we may conclude that $S_{\tg_{1}}(k)$, $S_{\tg_{2}}(k)$
have the same phases and the same pole structure, i.e., the graphs
$\tilde{\Gamma}_1$, $\tilde{\Gamma}_2$ are both isophasal and
isopolar. The graphs $\Gamma_1$, $\Gamma_2$ are therefore not only
isospectral, but also isoscattering (with respect to attaching leads
as in
$\tg_{1}=\nicefrac{\tilde{\Gamma}}{R_{1}},\,\tg_{2}=\nicefrac{\tilde{\Gamma}}{R_{2}}$).

We return to the example from section \ref{sec:A-basic-example}.
The quotients there are constructed according to the isospectral method
in \cite{BPB09} which guarantees the existence of a transplantation
$T$ the between graphs $\nicefrac{\tilde{\Gamma}}{\mathbf{1}_{H}}$
and $\nicefrac{\tilde{\Gamma}}{R}$. We note that both graphs are
composed of three leads, which form the so called {}``building blocks''
on which the transplantation acts:

\begin{equation}
\left(\begin{array}{c}
(Tf)|_{l_{1}}\\
(Tf)|_{l_{2}}\\
(Tf)|_{l_{3}}\end{array}\right)=\Pi\left(\begin{array}{c}
f|_{l_{1}}\\
f|_{l_{2}}\\
f|_{l_{3}}\end{array}\right),\quad\quad\Pi=\left(\begin{array}{ccc}
1 & 1 & 1\\
1 & -1 & 0\\
1 & 0 & -1\end{array}\right).\label{eq:tranp}\end{equation}
It can be easily checked that starting from any $f\in\Phi_{\nicefrac{\tilde{\Gamma}}{\mathbf{1}_{H}}}(k)$
and applying $T$ the function $Tf$ obeys the vertex condition of
$\nicefrac{\tilde{\Gamma}}{R}$ and therefore $Tf\in\Phi_{\nicefrac{\tilde{\Gamma}}{R}}(k)$.
In addition $T$ is invertible as can be seen from the matrix above.

We now treat the scattering properties of $\nicefrac{\tilde{\Gamma}}{\mathbf{1}_{H}}$
and $\nicefrac{\tilde{\Gamma}}{R}$. Using (\ref{eq:scatmat1H}),
(\ref{eq:scatmathR2d}) and (\ref{eq:tranp}) we confirm the validity
of (\ref{eq:scattering_of_isospectral}) for our example:

\begin{equation}
\Pi^{-1}S_{\nicefrac{\tilde{\Gamma}}{R}}(k)\Pi=S_{\nicefrac{\tilde{\Gamma}}{\mathbf{1}_{H}}}(k).\end{equation}

\section{More examples \label{sec:examples}}

\subsection{Various extensions to scattering systems}

\begin{figure}[H]
\includegraphics[scale=0.3]{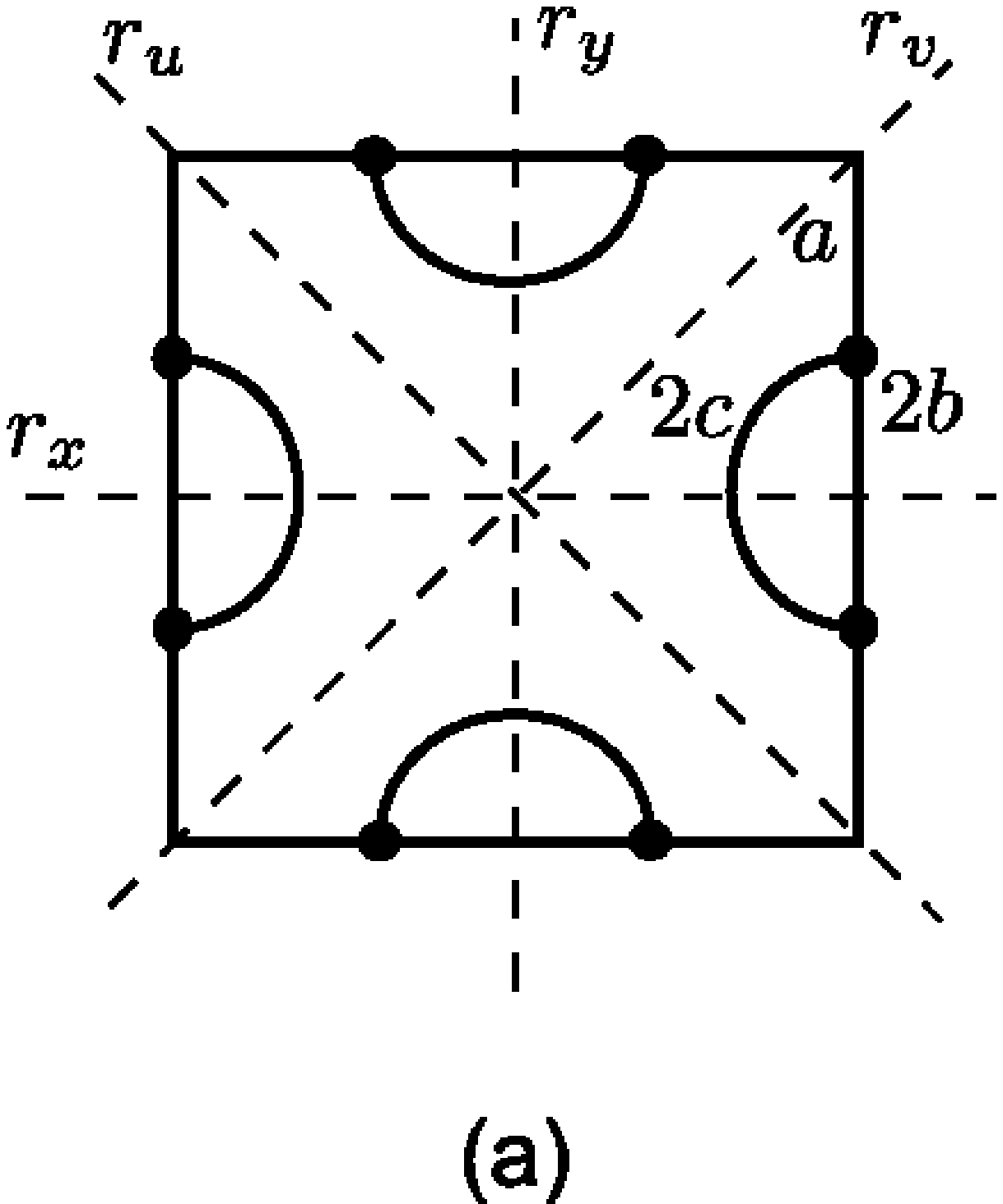}\qquad{}\qquad{}\qquad{}\qquad{}\qquad{}\includegraphics[scale=0.3]{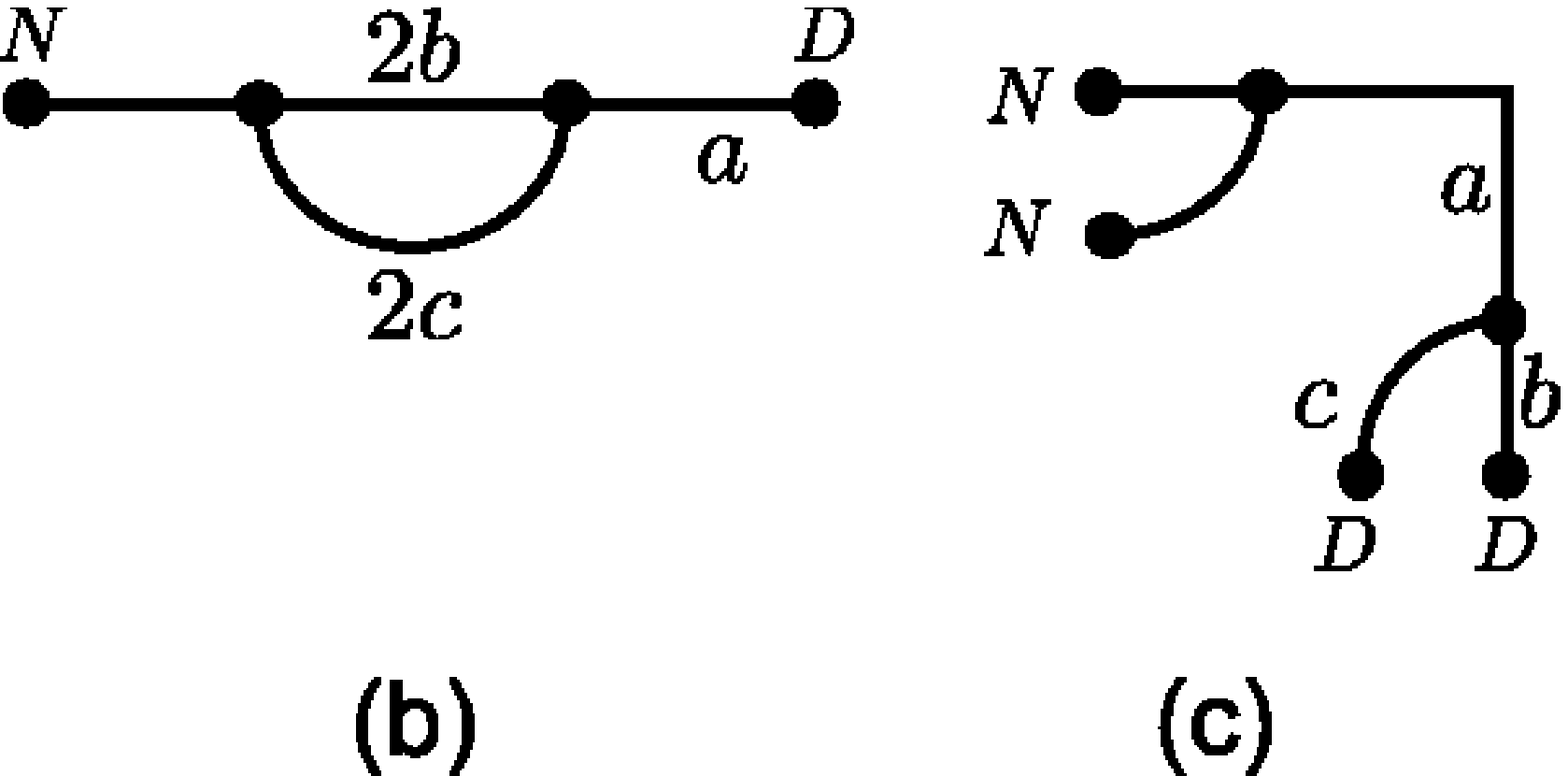}

\caption{(a) The graph $\Gamma$ that obeys the dihedral symmetry of the square
$D_{4}$. The lengths of some edges and the axes of the reflection
elements in $D_{4}$ are marked (b) The graph $\nicefrac{\Gamma}{R_{1}}$
(c) The graph $\nicefrac{\Gamma}{R_{2}}$. Neumann conditions are
default conditions in every unmarked vertex.}
\label{Flo:graphgamma}
\end{figure}

In this section we give further examples for the different ways to
extend isospectral graphs to isoscattering systems. Let $\Gamma$
be the quantum graph shown in figure \ref{Flo:graphgamma}(a). The
dihedral group $G=D_{4}$ is the symmetry group of this graph.

\begin{figure}[H]
\includegraphics[scale=0.3]{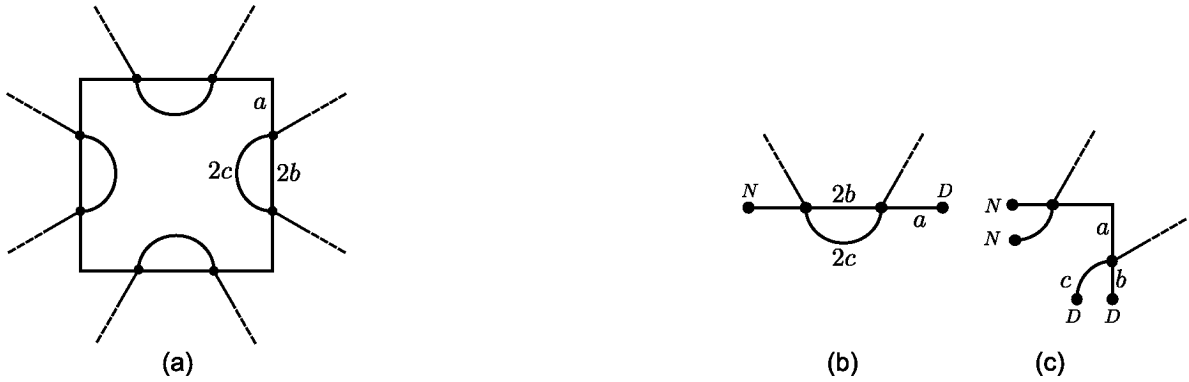}\qquad{}\qquad{}\qquad{}\qquad{}\qquad{}\qquad{}\includegraphics[scale=0.3]{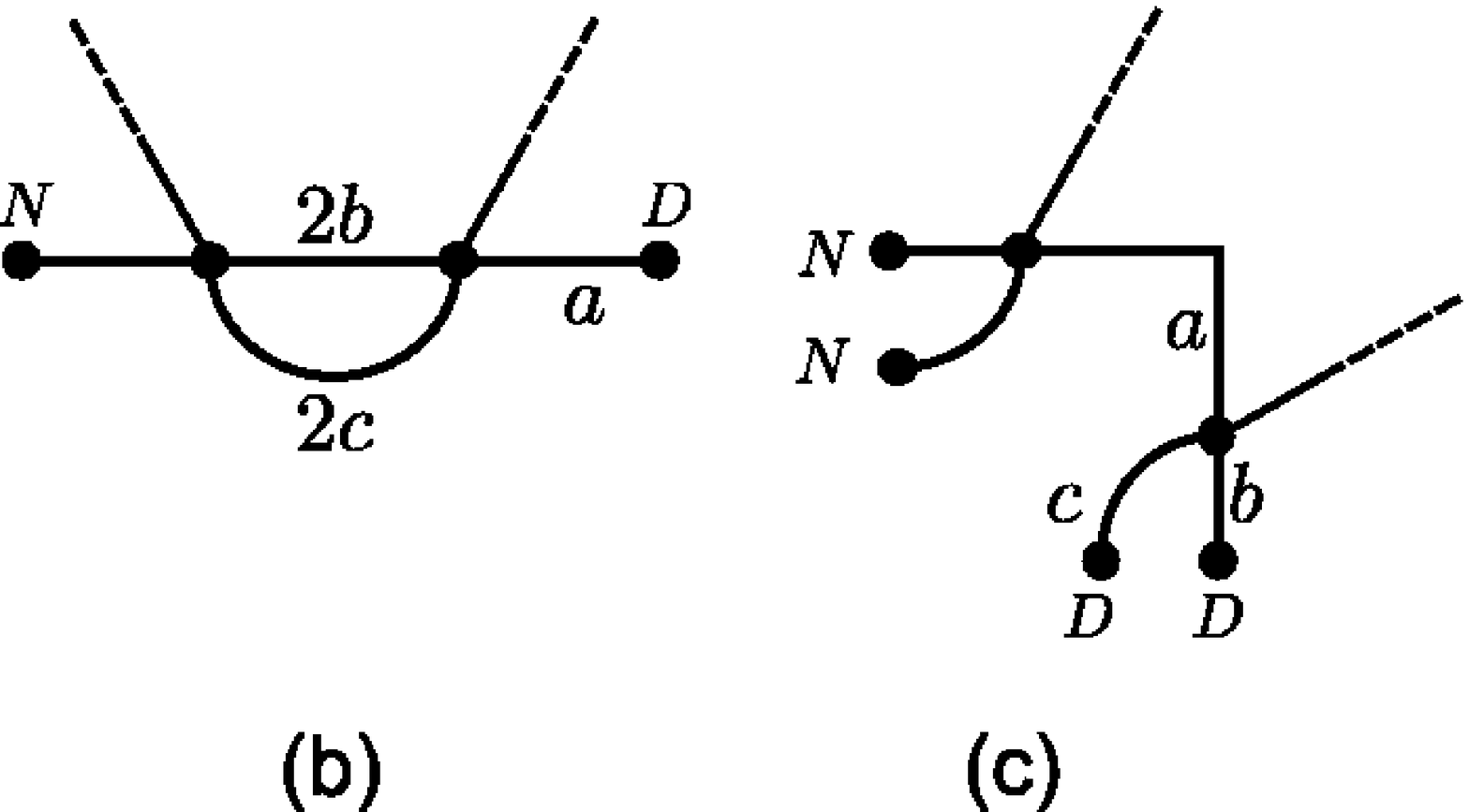}\caption{(a) The Graph $\tilde{\Gamma}$ with leads attached (b) The graph
$\nicefrac{\tilde{\Gamma}}{R_{1}}$ (c) The graph $\nicefrac{\tilde{\Gamma}}{R_{2}}$}
\label{Flo:exten}
\end{figure}

\begin{figure}[H]
\includegraphics[scale=0.3]{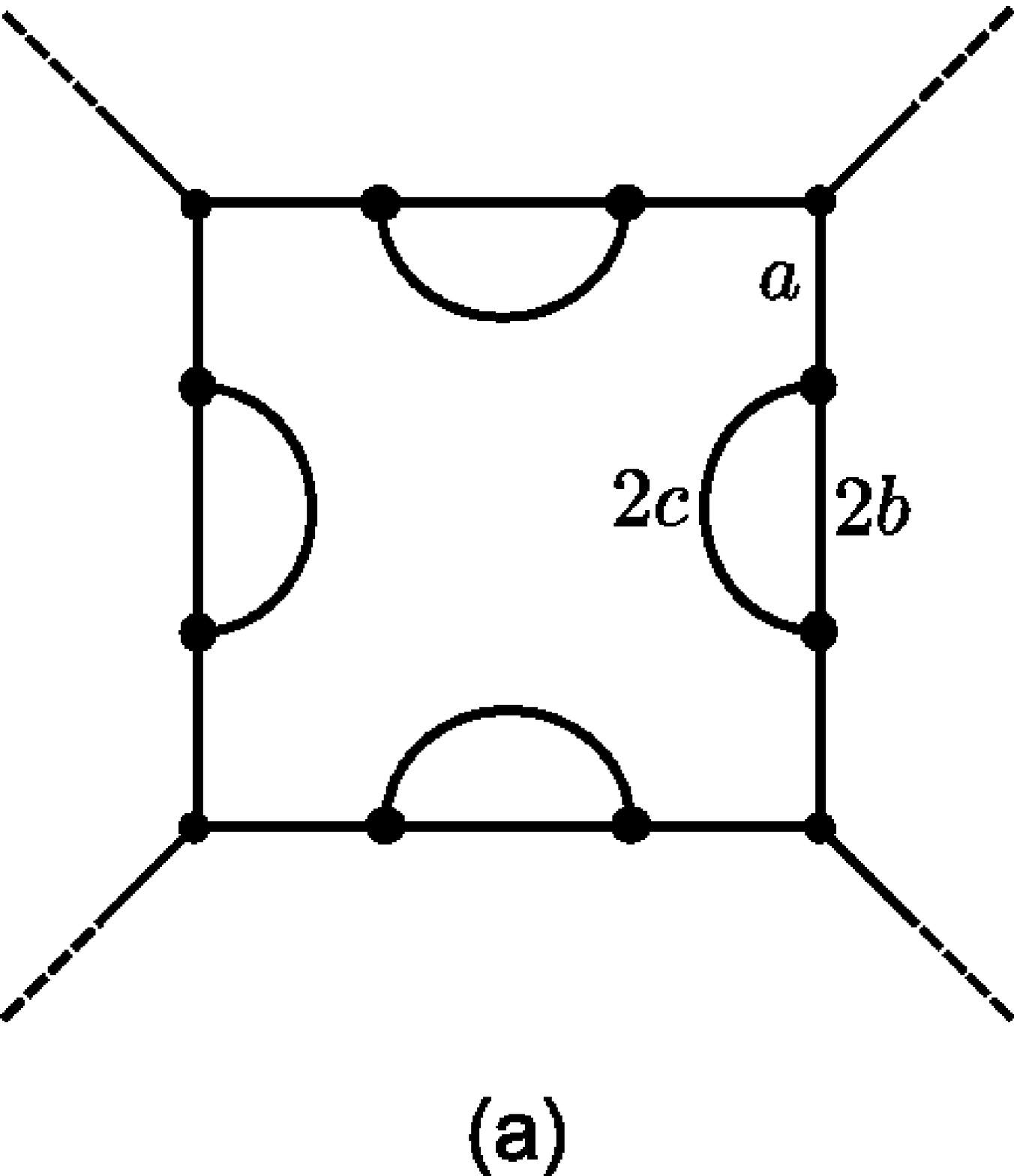}\qquad{}\qquad{}\qquad{}\qquad{}\includegraphics[scale=0.3]{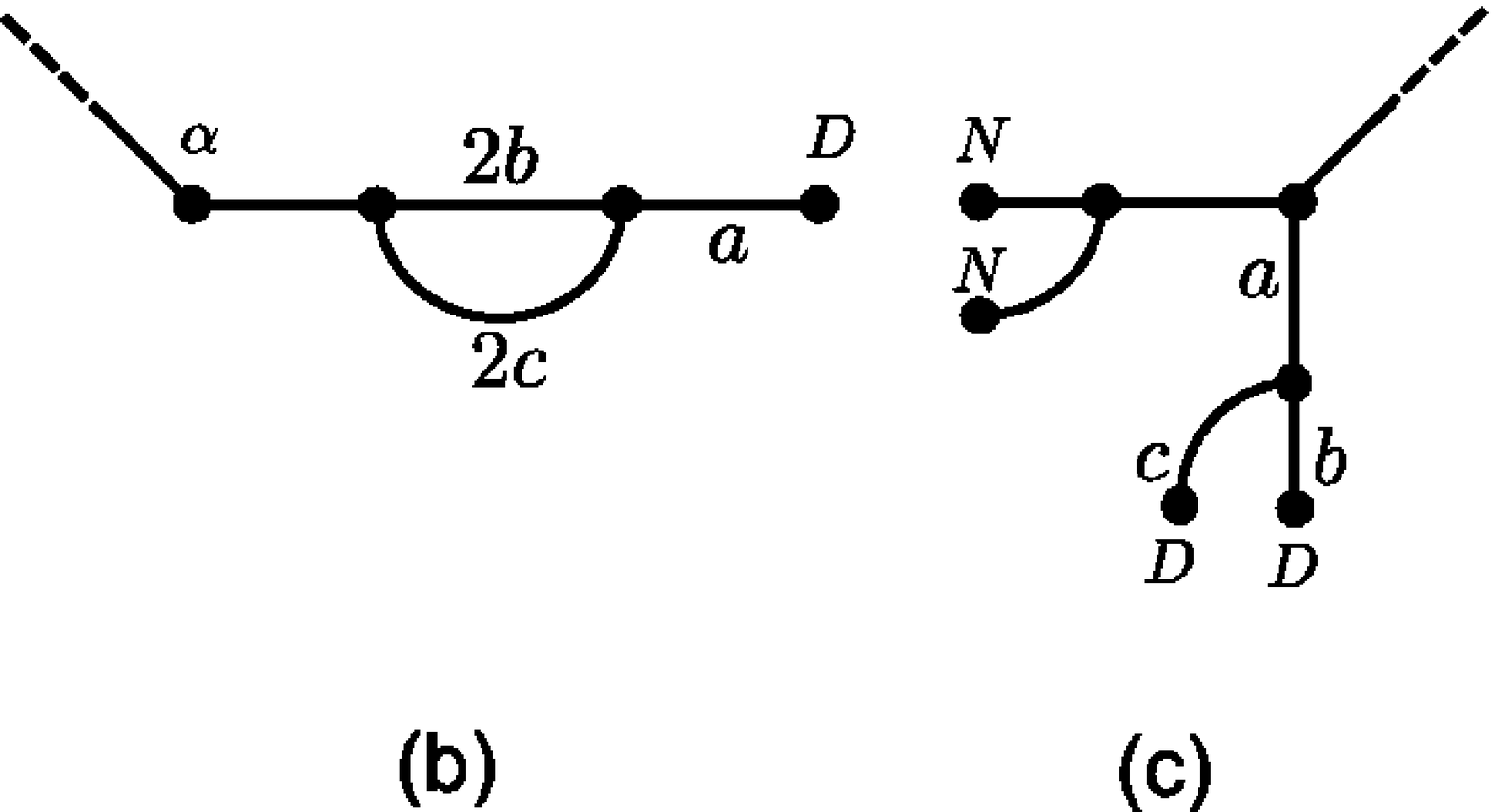}

\caption{(a) The Graph $\tilde{\Gamma}$ with leads attached (b) The graph
$\nicefrac{\tilde{\Gamma}}{R_{1}}$ (c) The graph $\nicefrac{\tilde{\Gamma}}{R_{2}}$}
\label{Flo:extquo}
\end{figure}

Examine the following subgroups of $G$

\[
H_{1}=\{e,\, r_{x},\, r_{y},\,\sigma^{2}\},\quad H_{2}=\{e,\, r_{u},\, r_{v},\,\sigma^{2}\},\]
where $r_{x},\, r_{y},\, r_{u},\, r_{v}$ denote reflections by the
axes $x,\, y,\, u,\, v$ and $\sigma$ is the counterclockwise rotation
by $\frac{\pi}{2}$. Consider the following one dimensional representations
$R_{1}$ and $R_{2}$ of $H_{1}$ and $H_{2}$ respectively

\begin{eqnarray*}
R_{1} & = & \{e\text{\textrightarrow}(1),\,\text{\ensuremath{\sigma}\textrightarrow}(\text{\textminus}1),\, r_{v}\text{\textrightarrow}(-1),\, r_{u}\text{\textrightarrow}(1)\},\\
R_{2} & = & \{e\text{\textrightarrow}(1),\text{\,\ensuremath{\sigma}\textrightarrow}(\text{\textminus}1),\, r_{y}\text{\textrightarrow}(1),\, r_{x}\text{\textrightarrow}(\text{\textminus}1)\}.\end{eqnarray*}
We have $\mathrm{Ind}_{H_{1}}^{G}R_{1}=\mathrm{Ind}_{H_{2}}^{G}R_{2}$
and therefore may use isospectral construction to obtain the two isospectral
quotient graphs $\nicefrac{\tilde{\Gamma}}{R_{1}}$ and $\nicefrac{\tilde{\Gamma}}{R_{2}}$
shown in figure \ref{Flo:graphgamma}(b),(c). We now consider two
possible extensions of $\Gamma$ to $\tilde{\Gamma}$ by adding infinite
leads (figures \ref{Flo:exten}(a), \ref{Flo:extquo}(a)). The action
of $G$ on the graph in figure \ref{Flo:exten}(a) is a free action.
This causes the vertices to which the leads are attached keep their
original (Neumann) vertex conditions from $\Gamma$. The isoscattering
quotients in this case appear in figure \ref{Flo:exten}(b),(c). The
case is different for the graph shown in figure \ref{Flo:extquo}(a)
in which the action of $r_{u}$, $r_{v}$ on the leads is not free.
This causes the disappearance of one of the leads in the quotient
$\nicefrac{\tilde{\Gamma}}{R_{2}}$ (figure \ref{Flo:extquo}(b)).
The surviving lead is attached to a vertex with the following vertex
conditions: \[
f_{e}(\alpha)=f_{l}(\alpha)\quad2f_{e}^{\prime}(\alpha)=f_{l}^{\prime}(\alpha),\]
where $e$ stands for edge and $l$ stands for lead. The disappearance
of the lead and the modified vertex conditions are typical for a quotient
derived from a non-free action (see \cite{BPB09,PB09} for a more
detailed description). The other quotient, $\nicefrac{\tilde{\Gamma}}{R_{1}}$
has Neumann conditions at the vertex where the lead is attached since
it is a quotient with respect to a representation of a group which
acts freely on $\tilde{\Gamma}$.

\subsection{The lack of transplantability\label{sec:lack_of_transplantability}}

\begin{figure}[h]
\includegraphics[scale=0.25]{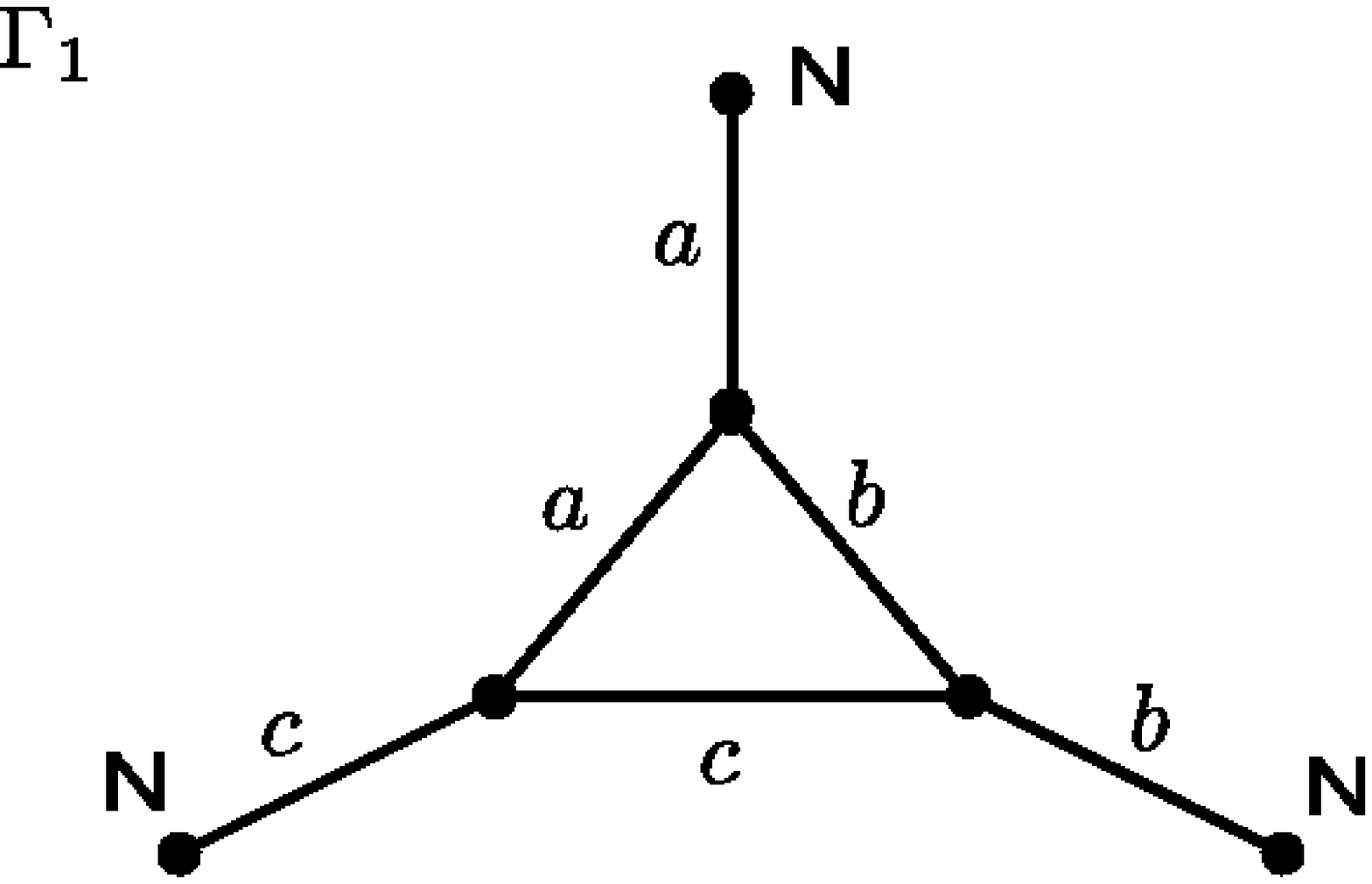}\qquad{}\qquad{}\qquad{}\qquad{}\includegraphics[scale=0.25]{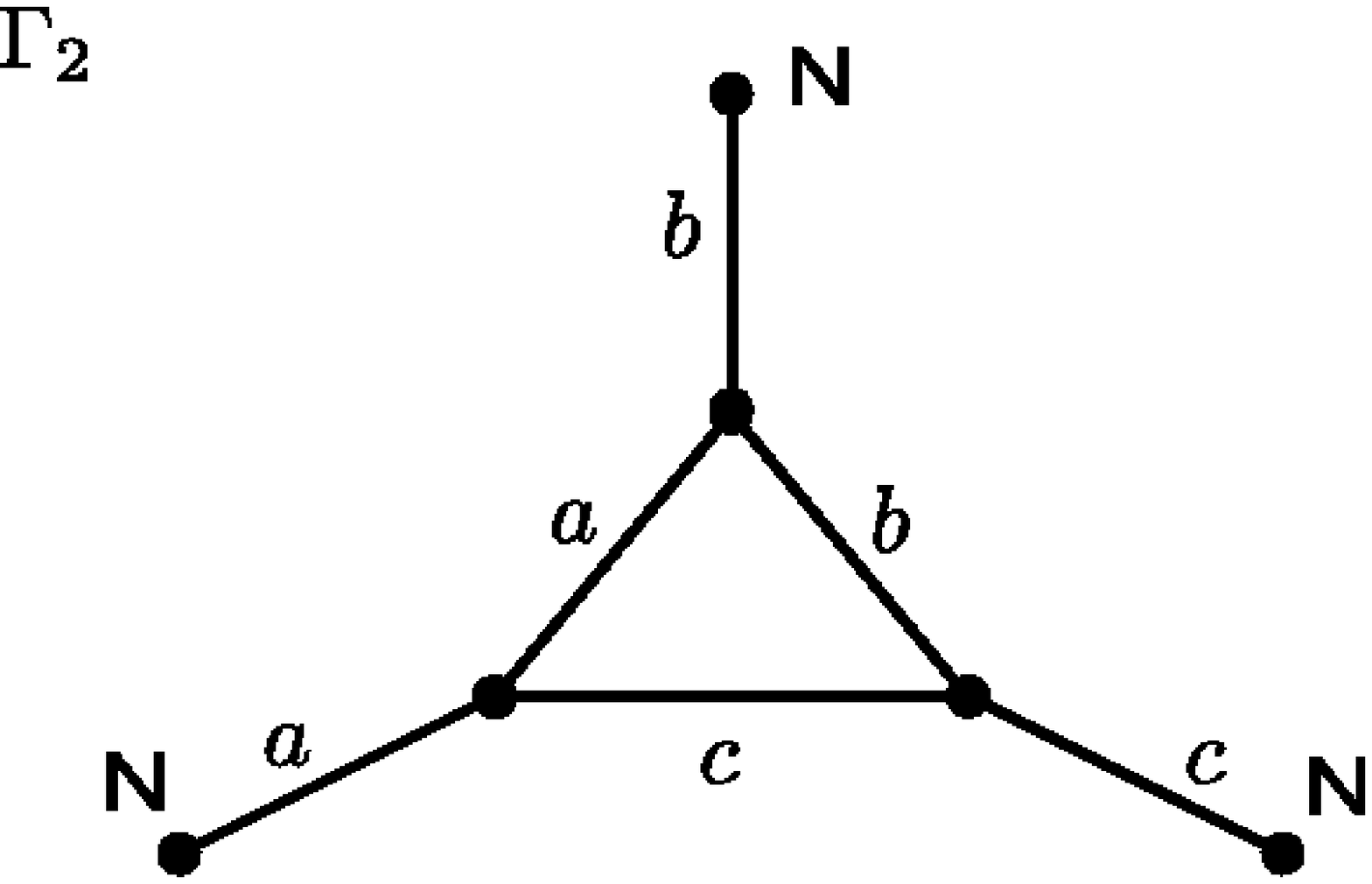}

\caption{The isospectral graphs $\Gamma_{1}$ and $\Gamma_{2}$. The lengths
of the edges are indicated. All vertices obey Neumann vertex conditions.}
\label{Flo:Idan12}
\end{figure}

\begin{figure}[H]
\includegraphics[scale=0.25]{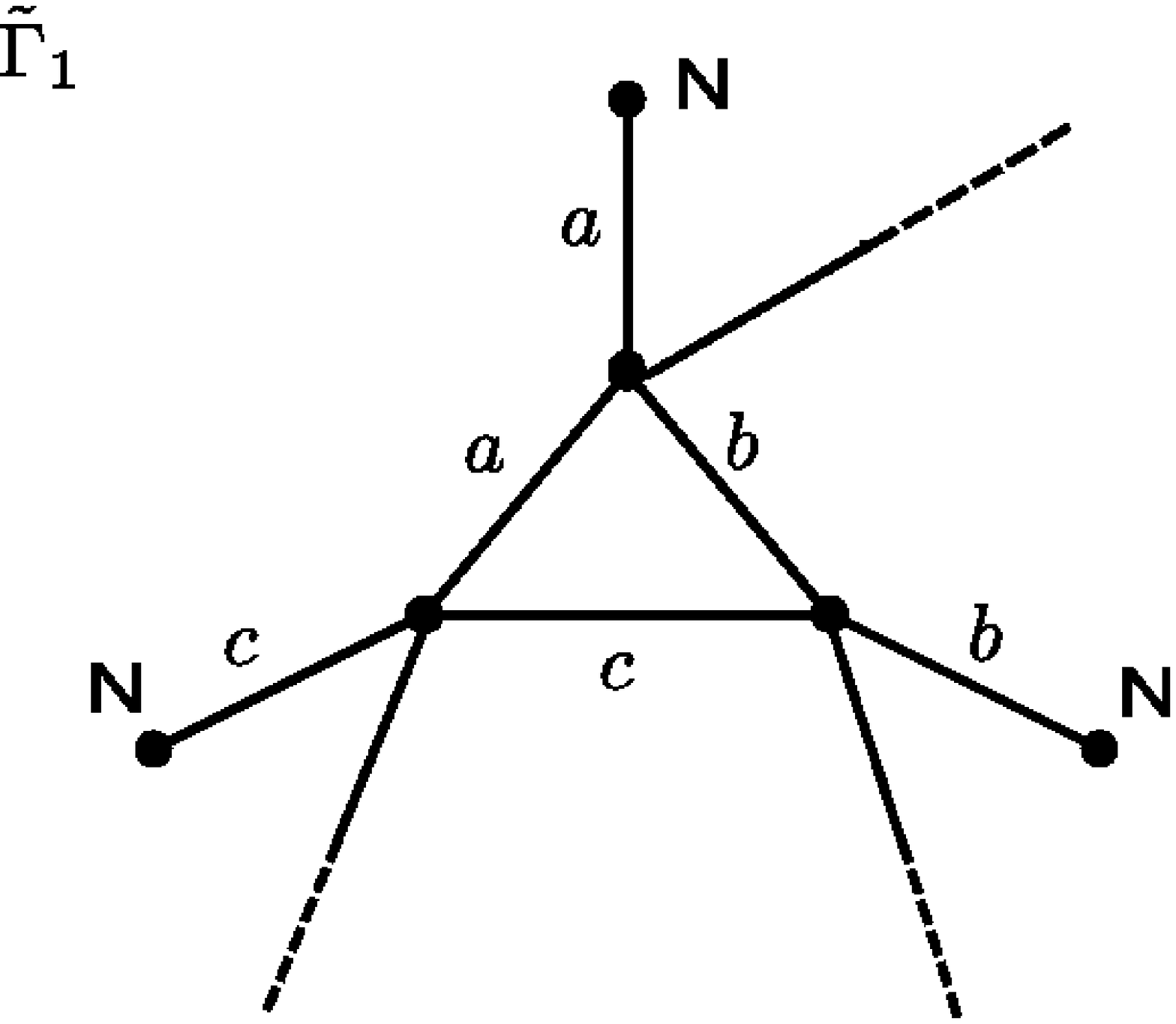}\qquad{}\qquad{}\qquad{}\qquad{}\includegraphics[scale=0.25]{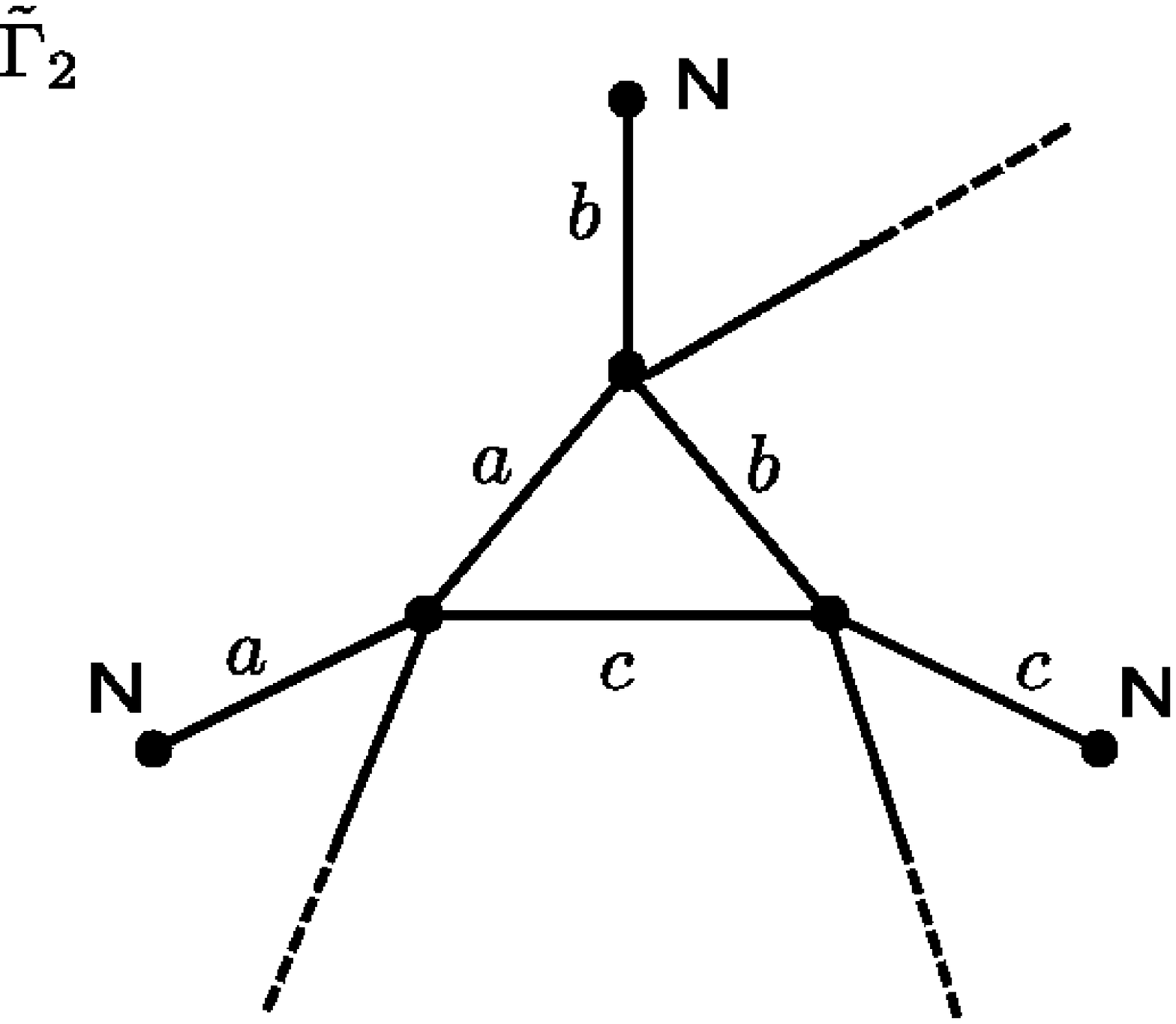}

\caption{The quantum graphs $\tilde{\Gamma}_{1}$ and $\tilde{\Gamma}_{2}$
which are extensions of $\Gamma_{1}$and $\Gamma_{2}$.}
\label{Flo:indanlead}
\end{figure}

Not all isospectral graph pairs are constructed by the Sunada method
and its generalizations. Here we discuss a specific example and consider
its scattering analogue. The isospectral quantum graphs shown in figure
\ref{Flo:Idan12} were constructed out of weighted discrete graphs.
The corresponding isospectral discrete graphs appeared in \cite{McDonaldMeyers03}
and were turned into quantum graphs in \cite{Oren08}, where their
isospectrality was proved by explicitly calculating the spectral determinants
and showing their equality. %
These graphs can be extended to form scattering systems by attaching
leads to their non-trivial vertices (figure \ref{Flo:indanlead}).

\begin{figure}[H]
\includegraphics[scale=0.1]{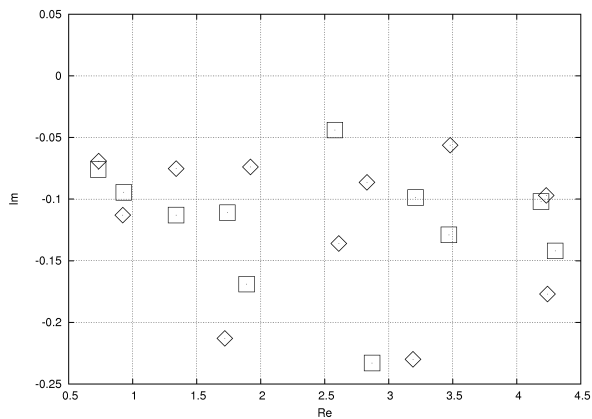}

\caption{Scattering resonances for graphs $\tilde{\Gamma}_{1}-\diamond$ and
$\tilde{\Gamma}_{2}-\square$.}
\label{Flo:poles}
\end{figure}

The corresponding scattering matrices were calculated and their poles
were numerically computed and are shown in figure \ref{Flo:poles}.
The different poles distributions indicate that the scattering matrices
are not conjugate, as opposed to the result obtained in section \ref{sec:isoscattering_graphs}.
We therefore conclude that there is no transplantation which relates
the values of the non-trivial vertices of these graphs.

.

\section{Discussion and open questions}

This paper discusses the linkage between isospectrality and isoscattering
of quantum graphs. We have described how to produce isoscattering
graphs using the recently developed isospectral construction method
\cite{BPB09,PB09}. Isoscattering graphs can be produced in two ways.
One can start from a certain graph with leads which forms a scattering
system, and construct out of it two graphs with leads that are isoscattering.
Another approach is to start from two graphs that are known to be
isospectral and discover all the possible ways in which leads can
be attached to the graphs to turn them into isoscattering. Both ways
are applications of the isospectral construction method and as such
they indicate that turning a graph into a scattering system do not
reveal more information on the graph then is already given by its
spectrum. This is compatible with the exterior-interior duality that
relates the spectral information with the scattering data. It is interesting
to compare this result with the conjecture brought in a recent work
by Okada et. al. \cite{OSTH05} in which the scattering from the exterior
of isospectral domains in $\mathbb{R}^{2}$ is discussed. The authors
conjecture that the distributions of the poles of the two scattering
matrices distinguish between the two domains. In other words, in spite
of the fact that the two domains are isospectral, Okada et. al. suggest
that they are not isoscattering. This proposition is not proved, but
it is ushered by heuristic arguments based on calculation of the Fredholm
determinant, and augmented by numerical simulations. The simulations
are performed on the isospectral domains in $\mathbb{R}^{2}$ that
appear in \cite{BCDS94}. We have shown that two isospectral graphs
can be turned into isoscattering systems in more than one possible
way. However, not every way of attaching leads to the isospectral
graphs would make them isoscattering. The leads should be connected
in a way which reflects the underlying symmetry that was used for
the construction of the isospectral pair. This may suggest that the
result in \cite{OSTH05} is a consequence of the fact that the outside
scattering problem of the isospectral drums breaks the symmetry that
was used in the isospectral drums construction. In \cite{BCDS94} these drums are obtained from a group of isometries of the hyperbolic plane. The construction there yields two quotients which are composed of seven copies of a hyperbolic triangle assembled in different configurations. The planar isospectral drums are then obtained by replacing
the fundamental hyperbolic triangle with a suitable Euclidean one. It should be noted that the hyperbolic drums are isometric and they become non-isometric only after the replacement with the Euclidean triangles. Therefore, the hyperbolic drums which obey the underlying symmetry are trivially isoscattering, being isometric.

The isoscattering property that we have obtained is a direct consequence
of the existence of a transplantation. The transplantation is guaranteed
by the isospectral construction method of \cite{BPB09,PB09}, but
also appears in other isospectral constructions. We therefore get
that the ability to turn isospectral graphs into isoscattering is
not restricted only to those graphs constructed by the discussed isospectral
method, but is possible for others as well, as long as the transplantation
exists. This observation supplies us with a technique to check the
transplantability property of two graphs that are known to be isospectral.
All we need to do is to connect leads to the sets of vertices which
we suspect to be transplantable and to check whether the corresponding
scattering matrices are conjugate. An example of an isospectral pair
which does not possess the transplantation property is given in section
\ref{sec:lack_of_transplantability}.

The key element of the isospectral theory developed in \cite{BPB09,PB09}
is the notion of the quotient graph. We have made use of the quotient
construction in the present paper as well, in order to produce isoscattering
pairs. While doing so we obtained a relation between the scattering
matrix of a graph and that of its quotient. This, not only sheds more
light on the properties of the quotient graph, but also gives an alternative
description of the quotient. As was already mentioned, the vertex
conditions of a quantum graph can be described in two ways. One is
by a linear set of equations on the function's values and derivatives
at the vertex. The other is by a scattering matrix which connects
the ingoing amplitudes to the outgoing ones at the vertex. The current
isospectral theory in \cite{BPB09,PB09} deals with vertex conditions
of the first type and describes the quotient construction in these
terms. Therefore, allowing to construct a quotient from a graph whose
vertex conditions are of the second type gives a new perspective to
the isospectral theory and broadens the class of graphs for which
the theory can be applied.

\section{Acknowledgments}

We acknowledge Idan Oren for suggesting us the new example of isospectral
graphs which is examined in section \ref{sec:lack_of_transplantability}.
We thank Akira Shudo and Ori Parzanchevski for fruitful discussions.
The work was supported by the Minerva Center for non-linear Physics,
the Einstein (Minerva) Center at the Weizmann Institute and the Wales
Institute of Mathematical and Computational Sciences (WIMCS). Grants
from EPSRC (grant EP/G021287), ISF (grant 166/09) and BSF (grant 2006065)
are acknowledged. The support by SFB/TR12 \textquoteleft{}Symmetries
and Universality in Mesoscopic Systems\textquoteright{} program of
the Deutsche Forschungsgemeischaft and Polish MNiSW grant no. DFG-SFB/38/2007
is gratefully acknowledged.

\section*{References}{}

\end{document}